# Multi-objective design of photon blockade for bright single-photon sources

Sunkyu Yu[1†], Xianji Piao[2§], and Namkyoo Park[3*]

[1]Intelligent Wave Systems Laboratory, Department of Electrical and Computer Engineering, Seoul National University, Seoul 08826, Korea

[2]Wave Engineering Laboratory, School of Electrical and Computer Engineering, University of Seoul, Seoul 02504, Korea

[3]Photonic Systems Laboratory, Department of Electrical and Computer Engineering, Seoul National University, Seoul 08826, Korea Seoul 08826, Korea

E-mail address for correspondence: sunkyu.yu@snu.ac.kr; piao@uos.ac.kr; nkpark@snu.ac.kr

**Abstract**

High-quality single-photon sources, realized through saturable emitters, photon blockade, or heralded pair generation, are indispensable building blocks for photonic quantum platforms. Although these mechanisms suppress multiphoton emission through distinct principles typically captured by analytical models, their practical implementation is constrained by conflicting requirements for purity, brightness, and indistinguishability, which must be balanced within high-dimensional design landscapes. Here, we propose a computational framework for optimizing competing metrics of single-photon sources. Building on a Liouville-space adjoint formulation that efficiently evaluates multiple objectives in Markovian open quantum systems, we develop a Jacobian-based update, which ensures first-order monotonic reduction of multi-objective costs. By incorporating simulated annealing to escape gradient-vanishing plateaus, our framework achieves a design success rate of nearly 60 % for photon blockade with $g_2(0) \leq 0.1$ and theoretically bounded brightness across a broad parameter space, without any analytical guidance. This framework provides a general recipe for multi-objective design of open quantum systems.

## Introduction

Achieving high-quality single-photon sources is an essential task for quantum technologies that rely on nonclassical light, such as quantum communication[1], photonic quantum computing[2], and quantum metrology[3]. A wide range of platforms has been explored, from quantum dots[4], colour centres[5], and trapped ions[6] to optical waveguides[7] and cavity quantum electrodynamics (QED) systems[8-13]. These platforms generate single photons through distinct mechanisms, including spontaneous parametric down conversion (SPDC)[7,14], saturable emission from driven quantum emitters[4-6], and Purcell-enhanced emission from an ensemble of ions[15]. Among them, photon blockade (PB)[8-13] offers a compact route to single-photon generation by exploiting optical nonlinearities to suppress multi-photon occupation, enabling antibunched emission in integrated photonic systems[16].

However, the design of high-quality single-photon sources remains a fundamental challenge because practical devices must balance multiple, often conflicting objectives across high-dimensional open-system parameter spaces. Key metrics, including single-photon purity, brightness, indistinguishability, spectral linewidth, and robustness[4,17], are not generally optimized simultaneously; for example, stronger pumping or outcoupling increases brightness but typically degrades antibunching or broadens the emission linewidth. This challenge is further exacerbated by the high-dimensional and generally nonconvex parameter-space landscapes of open quantum systems[18,19], which hinder stable convergence to an optimal design. Therefore, most approaches have relied on analytical few-mode models[8,12] along with intuition-guided engineering[9,10,13]. Because these approaches are often tailored to mechanism-specific models, they explore only restricted regions of the design space, limiting the discovery of unconventional operating regimes and hindering scalable optimization. In this context, a general framework for navigating such competing-objective landscapes in open quantum systems without relying on analytical guidance is highly desirable.

Here, we propose a problem-agnostic computational framework for optimizing single-photon sources. By formulating the adjoint method in a form generally applicable to Markovian open quantum systems, we efficiently evaluate multiple objectives associated with single-photon purity and brightness. To balance these competing objectives, we employ a Jacobian-descent framework that guarantees locally nonincreasing cost functions to first order. As an example, we apply the framework to photon blockade, where simulated annealing is used to overcome gradient vanishing near coherent states. The resulting approach achieves a design success rate of nearly 60 % for single-photon purity with $g_2(0) \leq 0.1$ and theoretically bounded brightness across a five-dimensional parameter space. The results further reveal a two-stage continuous transition from poor to high-quality single-photon emission in conventional photon blockade (CPB) without any analytical guidance. Our framework provides a general recipe for the inverse design of multi-objective open quantum systems across distinct mechanisms.

## Results

### Liouville-space adjoint method

Efficient evaluation of parameter-space gradients is central to the numerical design of target systems. Although the adjoint method[20,21] is well established in photonics as a standard tool for efficient gradient evaluation[22-26], its use in quantum physics has focused predominantly on time-dependent control[27-29], with a few exceptions[30,31]. To develop a cost-efficient optimization framework for single-photon sources characterized by steady states, we extend the adjoint method to $N$-dimensional Markovian open quantum systems described by the Lindblad quantum Master equation (QME)[32,33]:

$$\frac{d\rho}{dt} = -\frac{i}{\hbar}[H,\rho] + \sum_n d_n \left( L_n \rho L_n^\dagger - \frac{1}{2}\left\{ L_n^\dagger L_n, \rho \right\} \right) \equiv \mathcal{L}\rho, \tag{1}$$

where $\rho$ and $H$ are the density operator and Hamiltonian of the system, respectively, $\{L_n\}$ and $\{d_n\}$ denote jump operators and their corresponding decay rates, respectively, and the contribution from the Lamb shift is absorbed into $H$. In the Liouville-space representation[34], Eq. (1) is written as $d\rho_V/dt = L\rho_V$, where $L$ is the $N^2 \times N^2$ Liouville superoperator matrix and $\rho_V$ is the $N^2 \times 1$ column-vectorized density operator.

We focus on the optimization evaluated by a real-valued cost function $K(\mu, \lambda)$, where $\lambda \in \mathbb{R}^M$ denotes the physically accessible real-valued system parameters defining $L(\lambda)$, and $\mu$ denotes the column-vectorized steady-state density operator $\rho_S$. The optimization is subject to two constraints: the steady-state condition $L(\lambda)\mu = \mathbf{0}$ and the trace-normalization condition $I_V^\dagger\mu = 1$, where $I_V$ is the column-vectorized identity operator.

For gradient-based optimization, our objective is to compute the parameter-space gradient $D_\lambda K$, whose components are defined by $[D_\lambda K]_p = dK/d\lambda_p$. A direct application of the chain rule requires the computationally expensive steady-state sensitivities $J^T(\mu, \lambda)$ and $J^T(\mu^*, \lambda)$, where $J^T(a, b)$ denotes the transpose of the Jacobian matrix $J^T(a, b) = [\nabla_b a_1\ \nabla_b a_2 \dots \nabla_b a_N]$ with $[\nabla_b c]_p = \partial c/\partial b_p$ (Methods). The adjoint method avoids this computational cost by differentiating the steady-state ($L(\lambda)\mu = \mathbf{0}$) and trace-normalization ($I_V^\dagger\mu = 1$) constraints, together with their complex conjugates, and eliminating the explicit steady-state sensitivities. This process derives the parameter-space gradient, $D_\lambda K = \nabla_\lambda K + 2\mathrm{Re}[J^T(L\mu, \lambda)\varphi]$, where $\varphi$ denotes the adjoint state satisfying

$$\begin{bmatrix} L^T & I_V \\ \mu^T & 0 \end{bmatrix} \begin{bmatrix} \varphi \\ \nu \end{bmatrix} = \begin{bmatrix} -\nabla_\mu K \\ 0 \end{bmatrix}, \tag{2}$$

with the auxiliary multiplier $v$ (Methods).

Figure 1 illustrates the workflow of the adjoint method for optimizing general Markovian open quantum systems. Given an open quantum system specified by $L(\lambda)$ and a cost function $K$ (Fig. 1a), we calculate the steady state $\mu$ of $L(\lambda)$ (Fig. 1b). Using the gauge condition $\mu^{\mathrm{T}}\varphi = 0$ and the adjoint equation (Fig. 1c), we determine the adjoint state $\varphi$ with Eq. (2), which enables efficient evaluation of the parameter-space gradient of $K$ (Fig. 1d) for updating $L(\lambda)$.

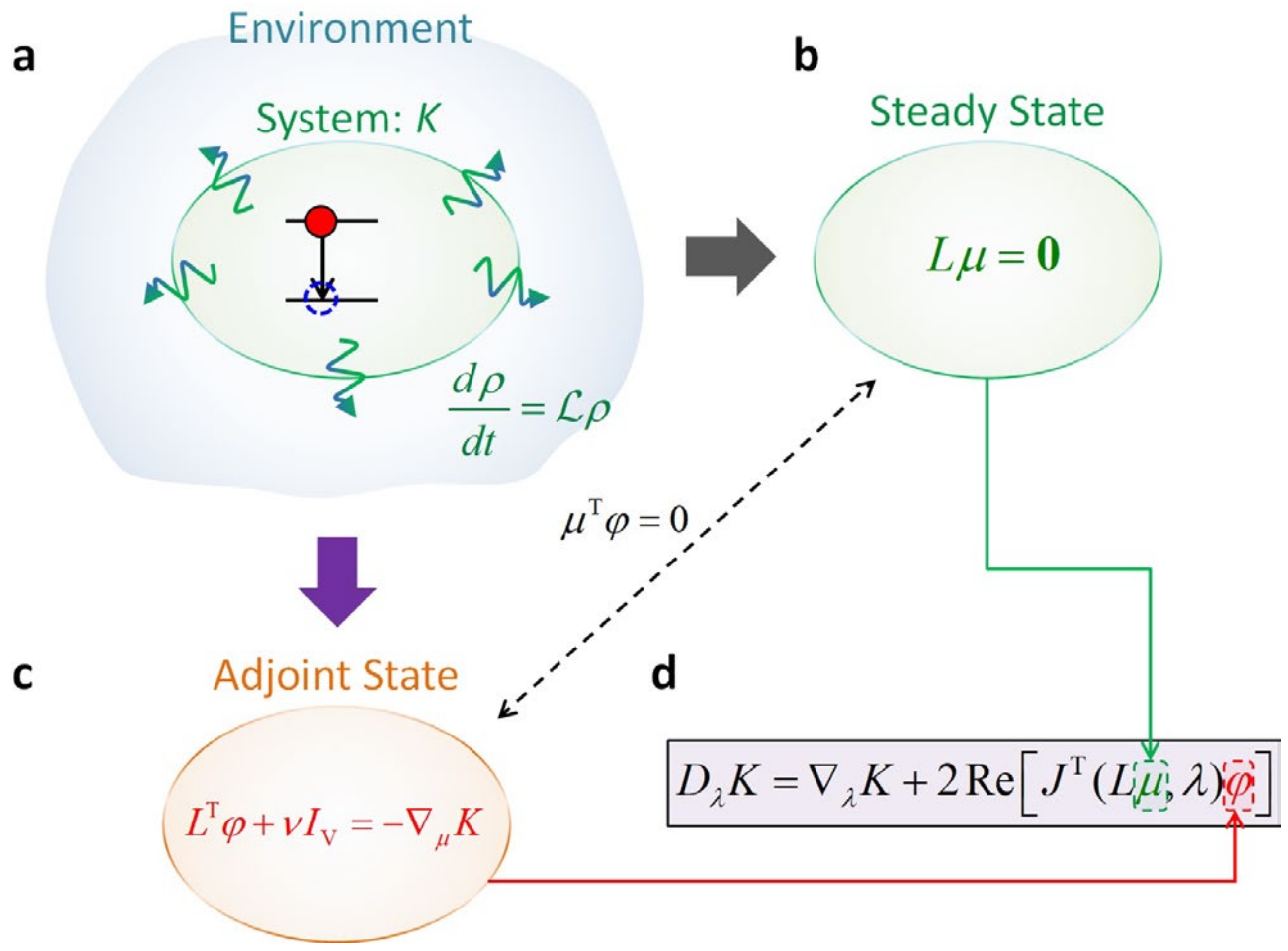

**Fig. 1. Liouville-space adjoint method**. **a,** Markovian open quantum system specified by $L(\lambda)$ and $K$. **b,** Steady-state calculation from $L(\lambda)\mu = 0$. **c,** Adjoint-state calculation under $\mu^{\mathrm{T}}\varphi = 0$. **d,** Parameter-space gradient evaluation using $\mu$, $\varphi$, and explicit parameter derivative of $L(\lambda)$.

## Cavity QED optimization

As an example of the proposed adjoint method, we investigate the optimization of a single-photon source in cavity QED. We consider a well-established scheme for CPB, governed by the Jaynes-Cummings (JC) Hamiltonian under an external coherent drive (Supplementary Note S1):

$$H = \hbar\left(\omega_{\mathrm{C}} - \omega\right)a^{\dagger}a + \left(\frac{E_{10} - \hbar\omega}{2}\right)\sigma_z - ig\left(a\sigma_+ - a^{\dagger}\sigma_-\right) + \xi\left(a + a^{\dagger}\right), \tag{3}$$

where $\hbar\omega_C$ and $\hbar\omega$ denote the cavity and driving photon energies, respectively, $E_{10}$ is the transition energy of a two-level system, referred to as the qubit, $g$ is the light-matter coupling energy, $\xi$ is the effective drive amplitude, $a$ and $a^\dagger$ are the cavity photon annihilation and creation operators, respectively, and $\sigma_{x,y,z}$ are the Pauli matrices. This Hamiltonian describes the interaction between a qubit and a cavity field (Fig. 2a), which can be represented as a Fock-state lattice[35] (Fig. 2b).

The open-system quantum dynamics in Eq. (1) further include three dominant dissipation processes (Fig. 2c): cavity relaxation, qubit relaxation, and qubit dephasing. Cavity relaxation is described by the jump operator $a$, comprising two loss mechanisms: output coupling to the single-photon collection channel at rate $\kappa_C$ and residual cavity loss with decay rate $\kappa_L$. Qubit relaxation and dephasing are described by the jump operators $\sigma_- = (\sigma_x - i\sigma_y)/2$ and $\sigma_z$, with the decay rates $\gamma$ and $\gamma_\varphi/2$, respectively.

Among the nine system parameters ($\omega_C$, $E_{10}$, $\omega$, $g$, $\xi$, $\kappa_C$, $\kappa_L$, $\gamma$, $\gamma_\varphi$) of the open JC model, we identify five practically tunable design parameters: $\omega_C$ and $\kappa_C$, controlled through the cavity design; $\omega$ and $\xi$, determined by the external drive; and $g$, engineered via the spatial overlap between the qubit and cavity field. Among the various performance figures for single photon sources, we focus on two target metrics[17]: the single-photon purity, quantified by minimizing $g_2(0) = \langle a^\dagger a^\dagger aa\rangle/\langle a^\dagger a\rangle^2$, and the brightness, defined as the photon outcoupling rate $B = \kappa_C\langle a^\dagger a\rangle$, where both quantities are evaluated in the steady state $\rho_S$.

To account for the broad dynamic ranges of the purity and brightness, we formulate the logarithmic cost functions, $K_{g2} = \ln(g_2(0) + g_0)$ and $K_B = -\ln(B + B_0)$, where $g_0$ and $B_0$ are small positive regularization constants that prevent logarithmic singularities. The optimization seeks a design parameter vector $\lambda = [\omega_C, \omega, g, \xi, \kappa_C]^T$ that simultaneously minimizes $K_{g2}$ and $K_B$ (Methods).

In the Fock-state picture, this multi-objective optimization corresponds to steering the steady-state population toward the single-photon manifold (dashed box in Fig. 2b).

Notably, this design problem without analytical intuition is highly challenging due to possibly conflicting objectives and the landscape containing broad plateaus and nonconvex structures. Figures 2d and 2e illustrate these effects through an example of the $g_2(0)$ and $B$ landscapes over ($g$, $\kappa_C$), respectively (Supplementary Note S2 for additional landscapes). For example, increasing $\kappa_C$ can enhance the output rate, but may also increase multiphoton emission (arrows in Figs. 2d,2e). Moreover, the wide plateau near $g_2(0) = 1$, referred to as the coherent plateau, and the presence of local minima (Supplementary Fig. S1), raise the risk of vanishing gradients and poor convergence.

To address the conflicting nature of $K_{g2}$ and $K_B$, a standard gradient-descent update $\lambda_{new} = \lambda - \beta u_G$, for example with $u_G = (1 - w)D_\lambda K_{g2} + wD_\lambda K_B$ ($\beta > 0$ and $0 \leq w \leq 1$), is insufficient, because $D_\lambda K_{g2}$ and $D_\lambda K_B$ are not aligned over substantial regions of the parameter space (arrows in Figs. 2d,2e). We therefore employ a Jacobian-descent update[36] (Fig. 2f), in which the objective gradients are assembled into the Jacobian matrix $J = [D_\lambda K_{g2}\ D_\lambda K_B]$. This matrix defines the dual cone, $C = \{u |\ J^T u \geq \mathbf{0}\}$, whose elements correspond to update directions for which a sufficiently small step along $-u$ is nonconflicting to first order[36]. We project each gradient vector onto $C$ by employing Moreau's decomposition[37] and a nonnegative least-squares solver[38], yielding the projected vectors, $P = [P_{g2}\ P_B]$. The update $\lambda_{new} = \lambda - \beta u_J$ with $u_J = (1 - w)P_{g2} + wP_B$ successfully guarantees locally nonincreasing cost functions to first order, in contrast to $u_G$. Because this guarantee requires a small step size, the learning rate $\beta$ is selected by a standard line search[39] at each epoch.

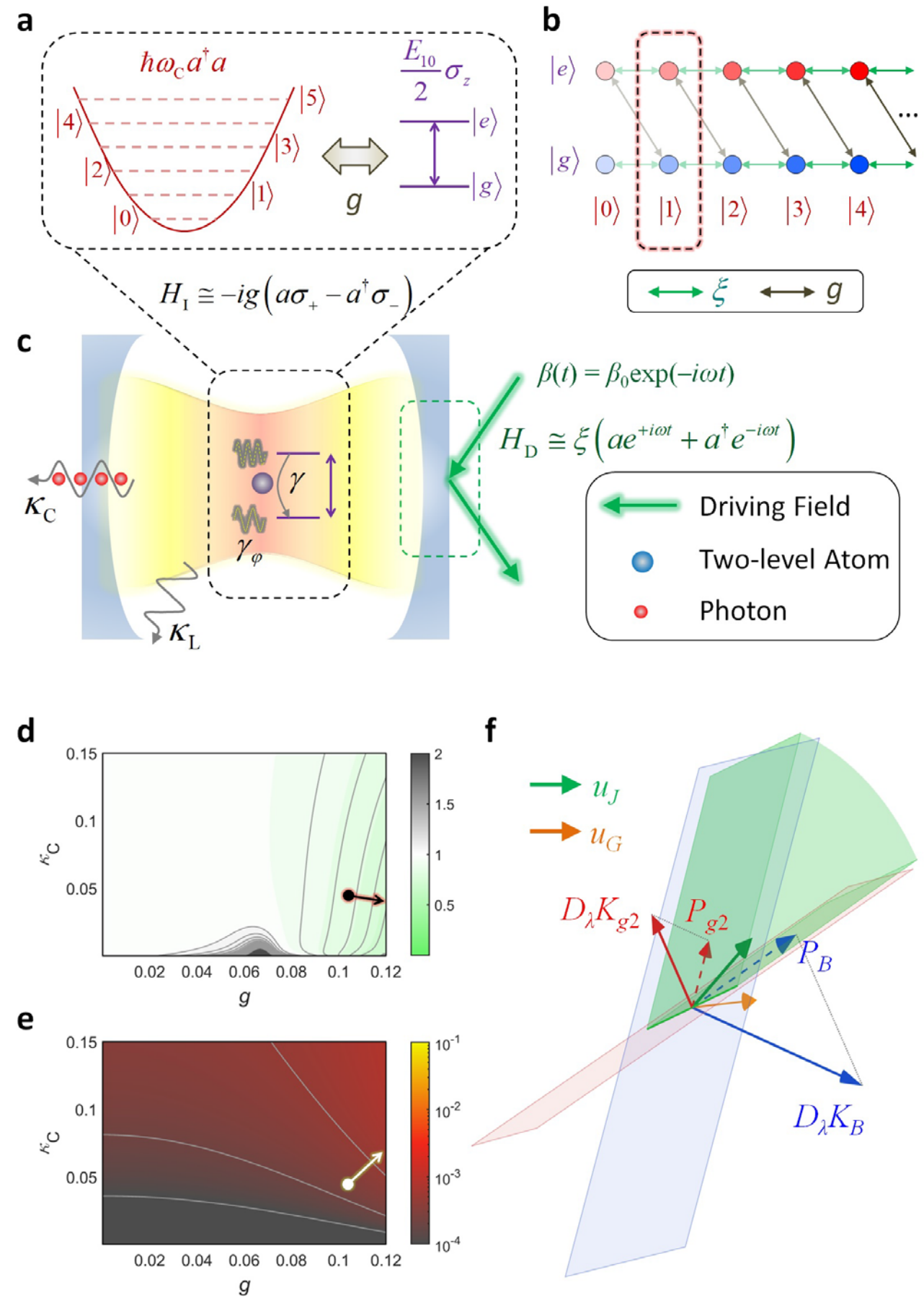

**Fig. 2. Multi-objective optimization of an open JC single-photon source**. **a,** JC coupling between a cavity mode and a qubit. **b,** Fock-state lattice representation. **c,** Cavity QED implementation. **d,e,** Landscapes of $g_2(0)$ (**d**) and $B$ (**e**) over ($g$, $\kappa_C$) at [$\omega_C$, $\omega$, $\xi$] = [0.974, 0.810, $8.7 \times 10^{-3}$], presenting conflicting local gradients with arrows. **f,** Nonconflicting update direction based on the Jacobian-descent update. Green region denotes the dual cone $C$. $w = 0.5$ in **f**.

### Transition to high-quality sources

To prioritize achieving the quantum characteristic essential for single-photon sources, our optimization policy first suppresses $K_{g2}$ when $g_2(0)$ exceeds a prescribed target threshold $g_{2,\mathrm{th}}$, by setting $u_G = D_\lambda K_{g2}$ with $w = 0$. Once this condition is satisfied with $g_2(0) \leq g_{2,\mathrm{th}}$, the projected Jacobian-descent direction is used to improve the brightness while preventing first-order degradation of the single-photon purity: $u_J = (1 - w)P_{g2} + wP_B$ with $w = \min[0.5, \max[0, g_{2,\mathrm{th}} -$ $g_2(0)$]]. While imposing physically motivated bounds on all design parameters to exclude invalid solutions (Methods), we examine an ensemble of 100 uniformly random realizations for each optimization condition.

Figure 3a shows representative optimization results under our policy for the target threshold $g_{2,\mathrm{th}} = 0.1$. Among the realizations, 33 % satisfy the design criterion—$g_2(0) \leq 0.1$ with finite $B$—following a continuous transition from the initial states to high-quality sources (Fig. 3b). By contrast, failed realizations exhibit two distinct behaviours: trapping at the coherent plateau near $g_2(0) = 1$, or excessive suppression of the brightness with $B = 0$, as shown in Fig. 3c (Supplementary Note S3 for other $g_{2,\mathrm{th}}$ values).

The successful trajectories reveal a two-stage transition toward high-quality sources (Fig. 3d). In stage I, the optimization decreases $g_2(0)$ by substantially reducing $\kappa_C$ and enhancing the CPB through increasing $g$. Unlike the relaxed-threshold case, such as $g_2(0) = 0.6$ (purple lines in Figs. 3e,3f), most realizations for stricter thresholds reach $B \approx 0$ during optimization (red lines in Figs. 3e,3f). Stage II uses the Jacobian-descent method to recover brightness by increasing $\kappa_C$ but substantially reducing $\xi$, thereby reviving $B$ while reducing, or at least preserving, the target single-photon purity (Supplementary Note S4).

Figure 3g summarizes the successful cases across various $g_2(0)$ thresholds, showing that the $g_2(0)$-$B$ performance is restricted by an analytical bound $B(g_{2,\mathrm{th}}) \sim \sqrt{g_{2,\mathrm{th}}}$, which is characterized by strong light-matter coupling and weak driving conditions (Supplementary Note S5). This result demonstrates that our method provides an optimal design near the theoretical bound, successfully identifying the continuous transition and its subprocesses, I and II, without any analytical guidance. However, the success probability remains as low as $P_S \approx 30$ % at the sub-Poisson thresholds of $g_2(0) < 1$ (Fig. 3h), indicating that escaping the coherent plateau remains a bottleneck in the optimization landscape.

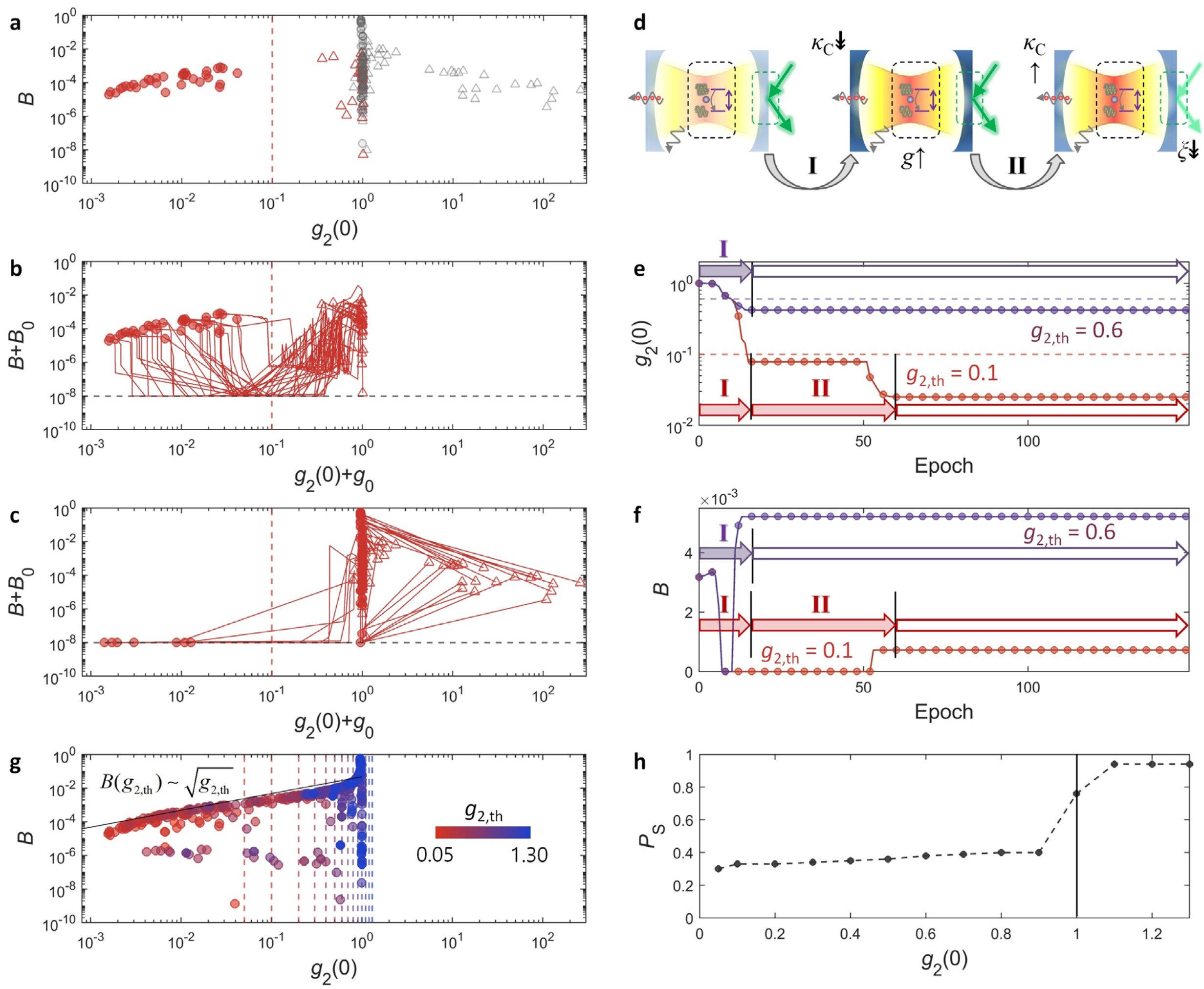

**Fig. 3. Continuous transition toward high-quality single-photon sources**. **a,** Initial (triangles) and final (circles) optimized sources for 100 realizations with the target threshold $g_2(0) = 0.1$ (red dashed line). Red and grey markers denote successful and failed realizations, respectively. **b,c,** Optimization trajectories for successful (**b**) and failed (**c**) realizations. **d,** Schematic of the two-stage transition mechanism. **e,f,** Evolutions of $g_2(0)$ (**e**) and $B$ (**f**) for $g_{2,\mathrm{th}} = 0.1$ (red) and 0.6 (purple). **g,** $g_2(0)$-$B$ distribution of successfully designed sources obtained for each $g_{2,\mathrm{th}}$. The black solid line denotes the analytical bound of brightness at a given threshold. **h,** Success probability $P_S$ versus the target threshold. The black solid line denotes the coherent plateau. Coloured dashed lines in **b,c,e,g** indicate the values of $g_{2,\mathrm{th}}$. Black dashed lines in **b,c** denote the value of $B_0$.

**Simulated annealing**

To alleviate trapping at the coherent plateau, we incorporate a simulated-annealing step[40] as a stochastic escape process (Fig. 4a; see Methods for details). At selected epochs, a trial design vector is generated by adding a temperature-dependent Gaussian perturbation to the current parameters, followed by projection onto the physically admissible parameter range. To determine the annealing step, we introduce an auxiliary energy function $E$ that favours low $g_2(0)$ and, once the purity threshold is satisfied, also rewards increasing $B$ (Methods). A proposed move is accepted if the energy decreases; otherwise, it is accepted with the Metropolis probability $\exp(-\Delta E/T_{SA})$, where $\Delta E$ is the energy difference between the trial and current states and $T_{SA}$ is the annealing temperature at the corresponding epoch. $T_{SA}$ is geometrically cooled from an initial value $T_0$ to a final value $T_{end}$, thereby progressively reducing the probability of accepting uphill moves. This stochastic step complements the Jacobian-descent update by enabling temporary escapes from shallow local minima and plateaus; the frequency of these escapes can be controlled by $T_0$ while keeping the other parameters, with larger $T_0$ promoting more frequent escapes.

Figures 4b and 4c show that the simulated-annealing-assisted trajectories cross the coherent plateau and reach the target region more frequently than the deterministic trajectories in Fig. 3b, thereby exploring regions of the parameter space that the latter fail to access (Supplementary Note S6). We further quantify this effect by scanning $T_0$ (Fig. 4d). Across a wide range of $T_0$, the success probability is consistently raised to approximately 60 %—roughly double the approximately 30 % obtained for the deterministic optimization without simulated annealing. Simulated annealing therefore operates as a controlled exploration mechanism that complements the Jacobian-descent update across the optimization landscape of the open quantum system.

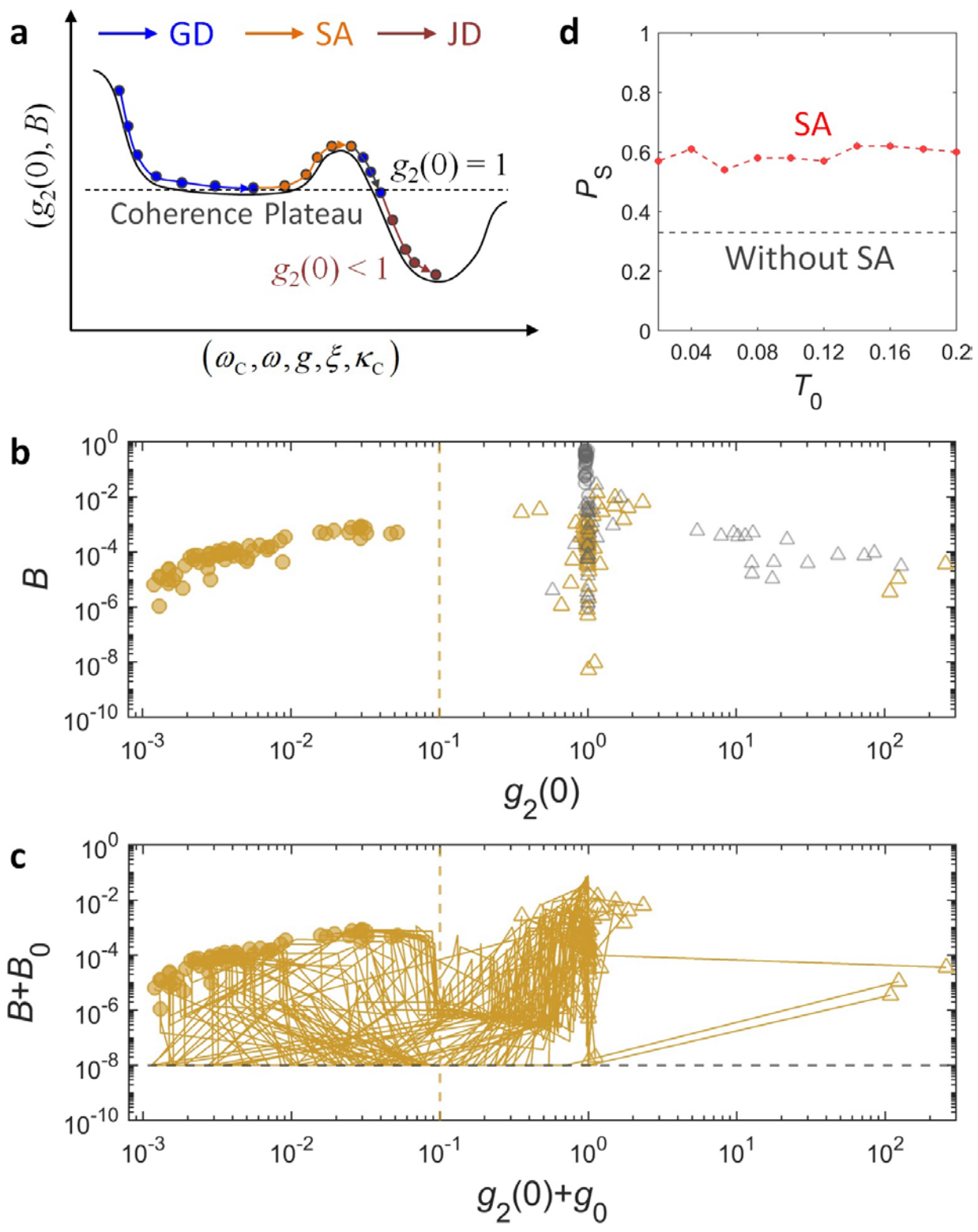

**Fig. 4. Simulated-annealing-assisted optimization. a,** Schematic of the optimization landscape, showing deterministic gradient descent (GD), stochastic simulated annealing (SA), and Jacobian descent (JD) across the coherent plateau near $g_2(0) = 1$. **b,** Initial (triangles) and final (circles) optimized sources for 100 realizations with the target threshold $g_2(0) = 0.1$ (red dashed line). Mustard and grey markers denote the successful and failed realizations. **c,** Optimization trajectories for successful realizations. $T_0 = 0.04$ in **b**,**c**. **d,** Success probability $P_S$ versus $T_0$, compared with the optimization (black dashed line) without simulated annealing.

## Discussion

The proposed method establishes a multi-objective adjoint-based optimization for the continuous transitions to high-quality single-photon sources under conflicting performance metrics. Although we focused on brightness and $g_2(0)$, the formulation is readily extensible to indistinguishability, spectral linewidth, and collection efficiency, by augmenting the objective Jacobian matrix. Moreover, our framework is not restricted to CPB. Because the method operates directly on the Lindbladian and user-defined cost functions, it can be extended to other forms of PB[10,12], SPDC[7,14], and saturable emission[4-6].

Continuous optimization trajectories obtained with our adjoint-method framework allow smooth tuning of $g_2(0)$ and $B$, rather than merely identifying isolated optima. Such a control is proper to task-dependent operation in reconfigurable elements[41], where the optimal purity-brightness trade-off can change with the operating condition, and to experimentally tunable single-photon sources[42], where performance must be adjusted in situ. Tracking the poor-to-high-quality transition therefore provides a practical route to reconfigurable quantum devices.

In the ultrastrong-coupling regime, the system-bath coupling must be expressed in the dressed eigenbasis[43,44]. Therefore, the jump operators and transition rates become dependent on the instantaneous eigenstates. Consequently, the parameter derivative of the Lindbladian must include not only the explicit derivatives of the Hamiltonian and decay rates, but also the derivatives of the dressed eigenstates, which will require particular care near quasi-degenerate dressed states.

In conclusion, we have developed a multi-objective adjoint optimization framework for Markovian open quantum systems and demonstrated its utility through the design of PB-based single-photon sources. The method enables the simultaneous optimization of conflicting source metrics approaching the theoretical bound and reveals continuous transition trajectories without any analytical guidance. These results establish a general recipe for scalable optimization in open quantum systems.

## Methods

**Derivation of the Liouville-space adjoint method.** For the cost function $K(\mu, \lambda) \in \mathbb{R}$, the chain rule derives its gradient $D_\lambda K$, as

$$D_\lambda K = \nabla_\lambda K + J^{\mathrm{T}}(\mu,\lambda)\left(\nabla_\mu K\right) + J^{\mathrm{T}}(\mu^*,\lambda)\left(\nabla_{\mu^*} K\right), \tag{4}$$

with $[J^{\mathrm{T}}(a, b)]_{pq} = \partial a_q/\partial b_p$. Note that whenever $(L\mu)$ appears as the first argument of $J^{\mathrm{T}}(a, b)$, the differentiation is taken as an explicit partial derivative in our notation. In contrast, $J^{\mathrm{T}}(\mu, \lambda)$ and $J^{\mathrm{T}}(\mu^*, \lambda)$ are quantities representing a total derivative, depicting the implicit $\lambda$-dependence of $\mu$ and $\mu^*$ through $L(\lambda)\mu = \mathbf{0}$. Therefore, $J^{\mathrm{T}}(\mu, \lambda)$ and $J^{\mathrm{T}}(\mu^*, \lambda)$ are computationally expensive to evaluate directly, because the differentiations require solving $L(\lambda)\mu = \mathbf{0}$ with $I_{\mathrm{V}}^\dagger\mu = 1$.

To circumvent this computational cost, we replace these terms using the derivatives obtained by differentiating $L(\lambda)\mu = \mathbf{0}$ and $I_{\mathrm{V}}^\dagger\mu = 1$, and their conjugates with respect to $\lambda$:

$$J^{\mathrm{T}}(\mu,\lambda)L^{\mathrm{T}} + J^{\mathrm{T}}(L\mu,\lambda) = O, \tag{5}$$

$$J^{\mathrm{T}}(\mu^*,\lambda)L^{\dagger} + J^{\mathrm{T}}(L^*\mu^*,\lambda) = O, \tag{6}$$

$$J^{\mathrm{T}}(\mu,\lambda)I_{\mathrm{V}} = \mathbf{0}. \tag{7}$$

$$J^{\mathrm{T}}(\mu^*,\lambda)I_{\mathrm{V}} = \mathbf{0}. \tag{8}$$

Note that while Eqs. (5) and (6) originate from $L(\lambda)\mu = \mathbf{0}$ including $Q$ equations, Eqs. (7) and (8) are obtained with $I_{\mathrm{V}}^\dagger\mu = 1$. Reflecting these degrees of freedom, we construct the Lagrangian using Eqs. (5)-(8) as constraints to Eq. (4), as follows:

$$\begin{aligned} D_\lambda K = {} & \nabla_\lambda K + J^{\mathrm{T}}(\mu,\lambda)\left(\nabla_\mu K\right) + J^{\mathrm{T}}(\mu^*,\lambda)\left(\nabla_{\mu^*} K\right) \\ & + \left[J^{\mathrm{T}}(\mu,\lambda)L^{\mathrm{T}} + J^{\mathrm{T}}(L\mu,\lambda)\right]\varphi + \left[J^{\mathrm{T}}(\mu,\lambda)I_{\mathrm{V}}\right]\nu \\ & + \left[J^{\mathrm{T}}(\mu^*,\lambda)L^{\dagger} + J^{\mathrm{T}}(L^*\mu^*,\lambda)\right]\varphi^* + \left[J^{\mathrm{T}}(\mu^*,\lambda)I_{\mathrm{V}}\right]\nu^*, \end{aligned} \tag{9}$$

where $\varphi \in \mathbb{C}^Q$ and $\nu \in \mathbb{C}$ are the Lagrange multipliers—referred to as the adjoint states—associated with the constraints in Eqs. (5)-(8), respectively. To eliminate the dependence on the high computational cost components, $J^{\mathrm{T}}(\mu, \lambda)$ and $J^{\mathrm{T}}(\mu^*, \lambda)$, we rearrange Eq. (9) as

$$D_\lambda K = \nabla_\lambda K + 2\,\mathrm{Re}\left[J^{\mathrm{T}}(L\mu,\lambda)\varphi\right] + 2\,\mathrm{Re}\left[J^{\mathrm{T}}(\mu,\lambda)\left(\nabla_\mu K + L^{\mathrm{T}}\varphi + I_{\mathrm{V}}\nu\right)\right], \tag{10}$$

which leads to the following adjoint equation:

$$L^{\mathrm{T}}\varphi + \nu I_{\mathrm{V}} = -\nabla_{\mu}K, \tag{11}$$

Equation (11) consists of $Q$ equations with $Q + 1$ unknowns: $\varphi \in \mathbb{C}^Q$ and $\nu \in \mathbb{C}$. The remaining degree of freedom corresponds to the gauge ambiguity, which we fix by imposing the trace-conservation gauge condition $\mu^{\mathrm{T}}\varphi = 0$, leading to the $(Q + 1) \times (Q + 1)$ matrix equation in Eq. (2), which is non-singular under the assumption of a unique steady state $\mu$. Using Eq. (2), we can uniquely determine the adjoint state $\varphi$ and the auxiliary multiplier $\nu$, which yield the gradient $D_\lambda K = \nabla_\lambda K + 2\mathrm{Re}[J^{\mathrm{T}}(L\mu, \lambda)\varphi]$. Therefore, the parameter-space gradient of $K$ is evaluated using $\mu$ and $\varphi$. The dominant computational cost—the forward and adjoint equation solves—is independent of the dimension $M$ of $\lambda$.

**Adjoint method for single-photon sources.** For the cost functions $K_{g2} = \ln(g_2(0) + g_0)$ and $K_B = -\ln(B + B_0)$ and the parameter vector $\lambda = [\omega_{\mathrm{C}}, \omega, g, \xi, \kappa_{\mathrm{C}}]^{\mathrm{T}}$, the associated gradients are

$$\nabla_{\lambda}K_{g2} = \mathbf{0}, \tag{12}$$

$$\nabla_{\lambda}K_{B} = -\frac{1}{B+B_0}\begin{bmatrix} 0 \\ 0 \\ 0 \\ 0 \\ \langle a^{\dagger}a\rangle_{\rho_{\mathrm{S}}} \end{bmatrix}. \tag{13}$$

Meanwhile, because $\langle O\rangle = \mathrm{Tr}[O\rho]$, its gradient with respect to the column-vectorized density operator $\mu \triangleq (\rho)_{\mathrm{V}}$ becomes $\nabla_\mu\langle O\rangle = (O^\dagger)_{\mathrm{V}}^*$, where $(\ldots)_{\mathrm{V}}$ denotes the column vectorization. Therefore, the density-vector gradients of the two cost functions are obtained as

$$\nabla_{\mu} K_{g2} = \frac{\left(a^{\dagger}a^{\dagger}aa\right)_{\mathrm{V}}^{*} - 2\frac{\left\langle a^{\dagger}a^{\dagger}aa\right\rangle_{\rho_{\mathrm{S}}}}{\left\langle a^{\dagger}a\right\rangle_{\rho_{\mathrm{S}}}}\left(a^{\dagger}a\right)_{\mathrm{V}}^{*}}{\left\langle a^{\dagger}a^{\dagger}aa\right\rangle_{\rho_{\mathrm{S}}} + g_0\left\langle a^{\dagger}a\right\rangle_{\rho_{\mathrm{S}}}^{2}}, \tag{14}$$

$$\nabla_{\mu} K_{B} = -\frac{\kappa_{\mathrm{C}}\left(a^{\dagger}a\right)_{\mathrm{V}}^{*}}{\kappa_{\mathrm{C}}\left\langle a^{\dagger}a\right\rangle_{\rho_{\mathrm{S}}} + B_0}. \tag{15}$$

We also need to evaluate the Jacobian $J^{\mathrm{T}}(L\mu, \lambda)$ along with Eqs. (12)-(15). For the $M$-dimensional real-valued parameter vector $\lambda = [\lambda_1, \lambda_2, \ldots, \lambda_M]^{\mathrm{T}}$, the Jacobian can be expressed as

$$J^{\mathrm{T}}(L\mu, \lambda) = \left[\frac{\partial L}{\partial \lambda_1}\mu \quad \frac{\partial L}{\partial \lambda_2}\mu \quad \cdots \quad \frac{\partial L}{\partial \lambda_M}\mu\right]^{\mathrm{T}}. \tag{16}$$

In evaluating $\partial L/\partial\lambda_k = \partial L_{\mathrm{H}}/\partial\lambda_k + \partial L_{\mathrm{L}}/\partial\lambda_k$, we utilize

$$\frac{\partial L_{\mathrm{H}}}{\partial \lambda_k} = -\frac{i}{\hbar}\left\{\left[I_N \otimes \frac{\partial H}{\partial \lambda_k}\right] - \left[\left(\frac{\partial H}{\partial \lambda_k}\right)^{\mathrm{T}} \otimes I_N\right]\right\}, \tag{17}$$

$$\frac{\partial L_{\mathrm{L}}}{\partial \lambda_k} = \left(\frac{\alpha}{\hbar}\right)^2 \frac{\partial}{\partial \lambda_k}\left[\sum_p d_p\left(L_p^{*} \otimes L_p - \frac{1}{2}\left(I_N \otimes L_p^{\dagger}L_p + L_p^{\mathrm{T}}L_p^{*} \otimes I_N\right)\right)\right]. \tag{18}$$

For the parameter vector $\lambda = [\omega_{\mathrm{C}}, \omega, g, \xi, \kappa_{\mathrm{C}}]^{\mathrm{T}}$ used in our example, the necessary derivatives for calculating the gradient are listed as follows:

$$\frac{\partial H}{\partial \lambda_1} = \hbar a^{\dagger}a, \tag{19}$$

$$\frac{\partial H}{\partial \lambda_2} = -\hbar a^{\dagger}a - \frac{\hbar}{2}\sigma_z, \tag{20}$$

$$\frac{\partial H}{\partial \lambda_3} = -i\left(a\sigma_{+} - a^{\dagger}\sigma_{-}\right), \tag{21}$$

$$\frac{\partial H}{\partial \lambda_4} = a + a^{\dagger}, \tag{22}$$

$$\frac{\partial L_{\mathrm{L}}}{\partial \lambda_5} = \left(\frac{\alpha}{\hbar}\right)^2 \left(a^* \otimes a - \frac{1}{2}\left(I_N \otimes a^\dagger a + a^{\mathrm{T}} a^* \otimes I_N\right)\right). \tag{23}$$

Using Eq. (16) together with the terms listed in Eqs. (17)-(23), Eq. (2) of the main text, and the gradients in Eqs. (12)-(15), we can evaluate $D_\lambda K = \nabla_\lambda K + 2\mathrm{Re}[J^{\mathrm{T}}(L\mu, \lambda)\varphi]$, which yields the gradient of each cost function, $K_{g2}$ and $K_B$.

**Design parameters.** For the parameter vector $\lambda = [\omega_{\mathrm{C}}, \omega, g, \xi, \kappa_{\mathrm{C}}]^{\mathrm{T}}$, the admissible domain is defined by $\omega - 0.2 \leq \omega_{\mathrm{C}} \leq \omega + 0.2$, $E_{10} - 0.2 \leq \omega \leq E_{10} + 0.2$, $0 \leq g \leq 0.12$, $10^{-4} \leq \xi \leq 0.2$, and $0 \leq \kappa_{\mathrm{C}} \leq 0.3$, with fixed parameters $E_{10} = \hbar = \alpha = 1$, $\kappa_{\mathrm{L}} = \gamma = \gamma_\varphi/2 = 0.001$, $g_0 = 10^{-5}$, and $B_0 = 10^{-8}$. The photon cutoff for Hilbert-space truncation is $N_{\mathrm{max}} = 6$.

**Simulated annealing.** At the $s$-th epoch, a trial parameter vector $\lambda'$ is generated by adding a Gaussian perturbation to the current vector $\lambda$, with the perturbation amplitude scaled by $(T_{\mathrm{SA}}/T_0)^{1/2}$, followed by projection onto the admissible parameter bounds. The temperature follows $T_{\mathrm{SA}}(s) = T_0(T_{\mathrm{end}}/T_0)^{(s-1)/(S-1)}$, where $S$ denotes the maximum epoch number. The trial is accepted according to the Metropolis rule using the auxiliary energy $E = \log_{10}(g_2(0) + g_0) - \alpha_B q \log_{10}(B + B_0)$, where $q = \min[1,\max(0,[g_{2,\mathrm{th}} - g_2(0)]/g_{2,\mathrm{th}})]$. Moves with $\Delta E \leq 0$ are always accepted, whereas uphill moves are accepted with probability $\exp(-\Delta E/T_{\mathrm{SA}})$. Accepted trials are used as the starting point for the subsequent deterministic line-search update. $\alpha_B = 0.2$ and $T_{\mathrm{end}} = 0.01$.

## Data availability

Data used in the current study are available from the corresponding authors upon request and can also be obtained by running the shared codes at 10.5281/zenodo.20742250 in the Zenodo.

## Code availability

All original code has been deposited at Zenodo and is publicly available at 10.5281/zenodo.20742250.

## Acknowledgements

We acknowledge financial support from the National Research Foundation of Korea (NRF) through the Basic Research Laboratory (No. RS-2024-00397664), Innovation Research Center (No. RS-2024-00413957), Core Research Grants (No. RS-2026-25469085), Pilot and Feasibility Grants (No. RS-2025-19912971), Young Researcher Program (No. RS-2025-00552989), and Midcareer Researcher Program (No. RS-2023-00274348), all funded by the Korean government. This work was supported by Creative-Pioneering Researchers Program through Seoul National University. We also acknowledge an administrative support from SOFT foundry institute.

## Author Contributions

All authors contributed equally to conceiving the idea, developing the code, discussing the results, and preparing the final manuscript.

## Competing Interests

The authors have no conflicts of interest to declare.

## Additional information

**Correspondence and requests for materials** should be addressed to S.Y., X.P., or N.P.

## References

1. Couteau, C., Barz, S., Durt, T., Gerrits, T., Huwer, J., Prevedel, R., Rarity, J., Shields, A. & Weihs, G. Applications of single photons to quantum communication and computing. *Nat. Rev. Phys.* **5**, 326-338 (2023).

2. Maring, N., Fyrillas, A., Pont, M., Ivanov, E., Stepanov, P., Margaria, N., Hease, W., Pishchagin, A., Lemaître, A. & Sagnes, I. A versatile single-photon-based quantum computing platform. *Nat. Photon.* **18**, 603-609 (2024).

3. Couteau, C., Barz, S., Durt, T., Gerrits, T., Huwer, J., Prevedel, R., Rarity, J., Shields, A. & Weihs, G. Applications of single photons in quantum metrology, biology and the foundations of quantum physics. *Nat. Rev. Phys.* **5**, 354-363 (2023).

4. Senellart, P., Solomon, G. & White, A. High-performance semiconductor quantum-dot single-photon sources. *Nat. Nanotechnol.* **12**, 1026-1039 (2017).

5. Neu, E., Steinmetz, D., Riedrich-Möller, J., Gsell, S., Fischer, M., Schreck, M. & Becher, C. Single photon emission from silicon-vacancy colour centres in chemical vapour deposition nano-diamonds on iridium. *New J. Phys.* **13**, 025012 (2011).

6. Walker, T., Miyanishi, K., Ikuta, R., Takahashi, H., Vartabi Kashanian, S., Tsujimoto, Y., Hayasaka, K., Yamamoto, T., Imoto, N. & Keller, M. Long-distance single photon transmission from a trapped ion via quantum frequency conversion. *Phys. Rev. Lett.* **120**, 203601 (2018).

7. Bock, M., Lenhard, A., Chunnilall, C. & Becher, C. Highly efficient heralded single-photon source for telecom wavelengths based on a PPLN waveguide. *Opt. Express* **24**, 23992-24001 (2016).

8. Imamoǧlu, A., Schmidt, H., Woods, G. & Deutsch, M. Strongly interacting photons in a

nonlinear cavity. *Phys. Rev. Lett.* **79**, 1467 (1997).

9. Birnbaum, K. M., Boca, A., Miller, R., Boozer, A. D., Northup, T. E. & Kimble, H. J. Photon blockade in an optical cavity with one trapped atom. *Nature* **436**, 87-90 (2005).

10. Snijders, H., Frey, J., Norman, J., Flayac, H., Savona, V., Gossard, A., Bowers, J., Van Exter, M., Bouwmeester, D. & Löffler, W. Observation of the unconventional photon blockade. *Phys. Rev. Lett.* **121**, 043601 (2018).

11. Chakram, S., He, K., Dixit, A. V., Oriani, A. E., Naik, R. K., Leung, N., Kwon, H., Ma, W.-L., Jiang, L. & Schuster, D. I. Multimode photon blockade. *Nat. Phys.* **18**, 879-884 (2022).

12. Zhou, Y.-H., Liu, T., Su, Q.-P., Zhang, X.-Y., Wu, Q.-C., Chen, D.-X., Shi, Z.-C., Shen, H. & Yang, C.-P. Universal photon blockade. *Phys. Rev. Lett.* **134**, 183601 (2025).

13. Hamsen, C., Tolazzi, K. N., Wilk, T. & Rempe, G. Two-photon blockade in an atom-driven cavity QED system. *Phys. Rev. Lett.* **118**, 133604 (2017).

14. Hong, C. & Mandel, L. Experimental realization of a localized one-photon state. *Phys. Rev. Lett.* **56**, 58 (1986).

15. Casabone, B., Deshmukh, C., Liu, S., Serrano, D., Ferrier, A., Hümmer, T., Goldner, P., Hunger, D. & de Riedmatten, H. Dynamic control of Purcell enhanced emission of erbium ions in nanoparticles. *Nat. Commun.* **12**, 3570 (2021).

16. Wang, J., Sciarrino, F., Laing, A. & Thompson, M. G. Integrated photonic quantum technologies. *Nat. Photon.* **14**, 273-284 (2020).

17. Buckley, S., Rivoire, K. & Vučković, J. Engineered quantum dot single-photon sources. *Reports on Progress in Physics* **75**, 126503 (2012).

18. Wu, R., Pechen, A., Rabitz, H., Hsieh, M. & Tsou, B. Control landscapes for observable preparation with open quantum systems. *J. Math. Phys.* **49** (2008).

19. Fentaw, H. W., Campbell, S. & Caton, S. Exploring quantum control landscape and solution space complexity through optimization algorithms and dimensionality reduction. *Sci. Rep.* **15**, 14605 (2025).

20. Givoli, D. A tutorial on the adjoint method for inverse problems. *Computer Methods in Applied Mechanics and Engineering* **380**, 113810 (2021).

21. Johnson, S. G. Notes on adjoint methods for 18.335. *Introduction to Numerical Methods* **732** (2012).

22. Wang, J., Shi, Y., Hughes, T., Zhao, Z. & Fan, S. Adjoint-based optimization of active nanophotonic devices. *Opt. Express* **26**, 3236-3248 (2018).

23. Hughes, T. W., Minkov, M., Williamson, I. A. & Fan, S. Adjoint method and inverse design for nonlinear nanophotonic devices. *ACS Photon.* **5**, 4781-4787 (2018).

24. Hughes, T. W., Minkov, M., Shi, Y. & Fan, S. Training of photonic neural networks through in situ backpropagation and gradient measurement. *Optica* **5**, 864-871 (2018).

25. Zhou, M., Liu, D., Belling, S. W., Cheng, H., Kats, M. A., Fan, S., Povinelli, M. L. & Yu, Z. Inverse design of metasurfaces based on coupled-mode theory and adjoint optimization. *ACS Photon.* **8**, 2265-2273 (2021).

26. Melo, E. G., Eshbaugh, W., Flagg, E. B. & Davanco, M. Multiobjective inverse design of solid-state quantum emitter single-photon sources. *ACS Photon.* **10**, 959-967 (2022).

27. Peirce, A. P., Dahleh, M. A. & Rabitz, H. Optimal control of quantum-mechanical systems: Existence, numerical approximation, and applications. *Phys. Rev. A* **37**, 4950-4964 (1988).

28. Khaneja, N., Reiss, T., Kehlet, C., Schulte-Herbrüggen, T. & Glaser, S. J. Optimal control of coupled spin dynamics: design of NMR pulse sequences by gradient ascent algorithms. *Journal of magnetic resonance* **172**, 296-305 (2005).

29. Petruhanov, V. N. & Pechen, A. N. GRAPE optimization for open quantum systems with time-dependent decoherence rates driven by coherent and incoherent controls. *J. Phys. A: Math. Theor.* **56**, 305303 (2023).

30. Vargas-Hernández, R. A., Chen, R. T., Jung, K. A. & Brumer, P. Fully differentiable optimization protocols for non-equilibrium steady states. *New J. Phys.* **23**, 123006 (2021).

31. Aydoğan, K., Schlimgen, A. W. & Head-Marsden, K. Stabilizing steady-state properties of open quantum systems with parameter engineering. *Phys. Rev. Res.* **7**, 023057, doi:10.1103/PhysRevResearch.7.023057 (2025).

32. Breuer, H.-P. & Petruccione, F. *The theory of open quantum systems* (Oxford University Press, 2002).

33. Manzano, D. A short introduction to the Lindblad master equation. *AIP Adv.* **10**, 025106 (2020).

34. Gyamfi, J. A. Fundamentals of quantum mechanics in Liouville space. *European Journal of Physics* **41**, 063002 (2020).

35. Saugmann, P. & Larson, J. Fock-state-lattice approach to quantum optics. *Phys. Rev. A* **108**, 033721 (2023).

36. Quinton, P. & Rey, V. Jacobian descent for multi-objective optimization. *arXiv preprint arXiv:2406.16232* (2024).

37. Moreau, J. J. Décomposition orthogonale d'un espace hilbertien selon deux cônes mutuellement polaires. *Comptes rendus hebdomadaires des séances de l'Académie des sciences* **255**, 238-240 (1962).

38. Lawson, C. L. & Hanson, R. J. *Solving least squares problems* (SIAM, 1995).

39. Nocedal, J. & Wright, S. J. *Numerical optimization* (Springer, 2006).

40. Van Laarhoven, P. J., Aarts, E. H., van Laarhoven, P. J. & Aarts, E. H. *Simulated Annealing: Theory and Applications* (Springer Dordrecht, 1987).

41. Valencia, N. H., Ma, A., Goel, S., Leedumrongwatthanakun, S., Graffitti, F., Fedrizzi, A., McCutcheon, W. & Malik, M. A large-scale reconfigurable multiplexed quantum photonic network. *Nat. Photon.* **20**, 202-207 (2026).

42. Nowak, A., Portalupi, S., Giesz, V., Gazzano, O., Dal Savio, C., Braun, P.-F., Karrai, K., Arnold, C., Lanco, L. & Sagnes, I. Deterministic and electrically tunable bright single-photon source. *Nat. Commun.* **5**, 3240 (2014).

43. Beaudoin, F., Gambetta, J. M. & Blais, A. Dissipation and ultrastrong coupling in circuit QED. *Phys. Rev. A* **84**, 043832 (2011).

44. Forn-Díaz, P., Lamata, L., Rico, E., Kono, J. & Solano, E. Ultrastrong coupling regimes of light-matter interaction. *Rev. Mod. Phys.* **91**, 025005 (2019).

## Figure Legends

**Fig. 1. Liouville-space adjoint method**. **a,** Markovian open quantum system specified by $L(\lambda)$ and $K$. **b,** Steady-state calculation from $L(\lambda)\mu = 0$. **c,** Adjoint-state calculation under $\mu^{\mathrm{T}}\varphi = 0$. **d,** Parameter-space gradient evaluation using $\mu$, $\varphi$, and explicit parameter derivative of $L(\lambda)$.

**Fig. 2. Multi-objective optimization of an open JC single-photon source**. **a,** JC coupling between a cavity mode and a qubit. **b,** Fock-state lattice representation. **c,** Cavity QED implementation. **d,e,** Landscapes of $g_2(0)$ (**d**) and $B$ (**e**) over ($g$, $\kappa_{\mathrm{C}}$) at [$\omega_{\mathrm{C}}$, $\omega$, $\xi$] = [0.974, 0.810, $8.7 \times 10^{-3}$], presenting conflicting local gradients with arrows. **f,** Nonconflicting update direction based on the Jacobian-descent update. Green region denotes the dual cone $C$. $w = 0.5$ in **f**.

**Fig. 3. Continuous transition toward high-quality single-photon sources**. **a,** Initial (triangles) and final (circles) optimized sources for 100 realizations with the target threshold $g_2(0) = 0.1$ (red dashed line). Red and grey markers denote successful and failed realizations, respectively. **b,c,** Optimization trajectories for successful (**b**) and failed (**c**) realizations. **d,** Schematic of the two-stage transition mechanism. **e,f,** Evolutions of $g_2(0)$ (**e**) and $B$ (**f**) for $g_{2,\mathrm{th}} = 0.1$ (red) and 0.6 (purple). **g,** $g_2(0)$-$B$ distribution of successfully designed sources obtained for each $g_{2,\mathrm{th}}$. The black solid line denotes the analytical bound of brightness at a given threshold. **h,** Success probability $P_{\mathrm{S}}$ versus the target threshold. The black solid line denotes the coherent plateau. Coloured dashed lines in **b,c,e,g** indicate the values of $g_{2,\mathrm{th}}$. Black dashed lines in **b,c** denote the value of $B_0$.

**Fig. 4. Simulated-annealing-assisted optimization.** **a,** Schematic of the optimization landscape, showing deterministic gradient descent (GD), stochastic simulated annealing (SA), and Jacobian descent (JD) across the coherent plateau near $g_2(0) = 1$. **b,** Initial (triangles) and final (circles) optimized sources for 100 realizations with the target threshold $g_2(0) = 0.1$ (red dashed line). Mustard and grey markers denote the successful and failed realizations. **c,** Optimization trajectories for successful realizations. $T_0 = 0.04$ in **b,c**. **d,** Success probability $P_{\mathrm{S}}$ versus $T_0$, compared with the optimization (black dashed line) without simulated annealing.

# Supplementary Information for "Multi-objective design of photon blockade for bright single-photon sources"

Sunkyu Yu[1†], Xianji Piao[2§], and Namkyoo Park[3*]

[1]Intelligent Wave Systems Laboratory, Department of Electrical and Computer Engineering, Seoul National University, Seoul 08826, Korea

[2]Wave Engineering Laboratory, School of Electrical and Computer Engineering, University of Seoul, Seoul 02504, Korea

[3]Photonic Systems Laboratory, Department of Electrical and Computer Engineering, Seoul National University, Seoul 08826, Korea

E-mail address for correspondence: [†]sunkyu.yu@snu.ac.kr, [§]piao@uos.ac.kr, [*]nkpark@snu.ac.kr

**Supplementary Note S1. Open quantum dynamics in the driven JC model**

For completeness, we provide a self-contained derivation of the open-system driven JC model in cavity QED used in the main text. We begin with the isolated quantum Rabi model (QRM) in the Coulomb gauge [1]:

$$H_{\mathrm{C}} = \hbar\omega_{\mathrm{C}} a^{\dagger} a + \frac{E_{10}}{2}\left[\sigma_z \cos\left(\frac{2A_0 d\left(a^{\dagger}+a\right)}{\hbar}\right) + \sigma_y \sin\left(\frac{2A_0 d\left(a^{\dagger}+a\right)}{\hbar}\right)\right], \tag{S1}$$

where $\hbar\omega_{\mathrm{C}}$ and $E_{10}$ denote the cavity photon energy and the two-level transition energy, respectively, $A_0$ is the vacuum amplitude of the vector potential, $d$ is the differential dipole moment, $a$ and $a^{\dagger}$ are the cavity photon annihilation and creation operators, respectively, and $\sigma_{x,y,z}$ are the Pauli matrices. By assuming the weak-coupling regime $\eta = A_0 d/\hbar \ll 1$ and neglecting the rapidly oscillating terms under the rotating-wave approximation (RWA), we obtain the JC model:

$$H_{\mathrm{JC}} = \hbar\omega_{\mathrm{C}} a^{\dagger} a + \frac{E_{10}}{2}\sigma_z - i\eta E_{10}\left(a\sigma_+ - a^{\dagger}\sigma_-\right). \tag{S2}$$

An external coherent drive is incorporated through the dipole interaction Hamiltonian:

$$H_{\mathrm{I}} = -\int_V \mathbf{P}_{\mathrm{ext}} \cdot \mathbf{E} dv = \varepsilon_{\mathrm{d}}\left(b - b^{\dagger}\right)\left(a - a^{\dagger}\right) \approx -\varepsilon_{\mathrm{d}}\left(ab^{\dagger} + a^{\dagger}b\right), \tag{S3}$$

where $b$ and $b^{\dagger}$ denote the annihilation and creation operators of the external bosonic mode, $\varepsilon_{\mathrm{d}}$ denotes the coupling term encompassing the volume integral, and the counter-rotating terms have been neglected under the RWA. Assuming that the external mode is prepared in a coherent state, we apply the $c$-number substitution, $b \rightarrow \beta(t) = \beta_0 \exp(-i\omega t)\exp(i\theta)$ ($\beta_0 \geq 0$), which leads to

$$H_{\mathrm{I}} = -\varepsilon_{\mathrm{d}}\left(\beta(t)^{*} a + \beta(t) a^{\dagger}\right) = \xi\left(a e^{+i\omega t} e^{-i\theta} + a^{\dagger} e^{-i\omega t} e^{+i\theta}\right), \tag{S4}$$

where $\xi = -\beta_0\varepsilon_{\mathrm{d}}$ denotes the corresponding effective drive amplitude. With this driving Hamiltonian, the governing equation for our study is the driven JC model:

$$H = \hbar\omega_C a^\dagger a + \frac{E_{10}}{2}\sigma_z - i\eta E_{10}\left(a\sigma_+ - a^\dagger\sigma_-\right) + \xi\left(ae^{+i\omega t}e^{-i\theta} + a^\dagger e^{-i\omega t}e^{+i\theta}\right). \qquad \text{(S5)}$$

By applying the rotating-frame transformation for the drive frequency $\omega$, we can utilize the following time-independent Hamiltonian for analysing open quantum dynamics:

$$H = \hbar\left(\omega_C - \omega\right)a^\dagger a + \left(\frac{E_{10} - \hbar\omega}{2}\right)\sigma_z - i\eta E_{10}\left(a\sigma_+ - a^\dagger\sigma_-\right) + \xi\left(ae^{-i\theta} + a^\dagger e^{+i\theta}\right). \qquad \text{(S6)}$$

Because the system contains only a single coherent drive, we set $\theta = 0$ without loss of generality, obtaining Eq. (3) in the main text, where the light-matter coupling energy is defined as $g \triangleq \eta E_{10}$. In our optimization, we target the experimentally tunable system parameters: cavity frequency $\omega_C$, driving frequency $\omega$, light-matter coupling strength $g$, and the driving amplitude $\xi$.

In terms of engineering the system environment, we consider the system-bath coupling in the Lindblad QME, as summarized in Table S1. Among the decay parameters ($\kappa_C$, $\kappa_L$, $\gamma$, $\gamma_\varphi$), we focus on engineering the practically tunable parameter, $\kappa_C$, which is associated with the cavity quality factor. Therefore, the full design parameter vector for optimization is $\lambda = [\omega_C, \omega, g, \xi, \kappa_C]^T$.

**Table S1. Jump operators and decay rates.** In the Lindblad QME, we include two relaxation processes—cavity and qubit relaxations—and one qubit dephasing process.

| | Cavity relaxation | Qubit relaxation | Qubit dephasing |
|---|---|---|---|
| Jump operators | $a$ | $\sigma_-$ | $\sigma_z$ |
| Decay rates | $\kappa_C + \kappa_L$ | $\gamma$ | $\gamma_\varphi/2$ |

## Supplementary Note S2. Parameter-space landscapes

Figure S1 shows additional examples of the optimization landscapes for different pairs of design parameters. As discussed in the main text, these landscapes contain broad regions with conflicting gradient directions, frequently exhibit coherent plateaus at $g_2(0) = 1$, and possess multiple local minima (Figs. S1a, S1b, S1e, and S1h). These observations demonstrate that the optimization of single-photon sources is challenging because of conflicting objectives, nonconvex landscape geometry, and vanishing gradients.

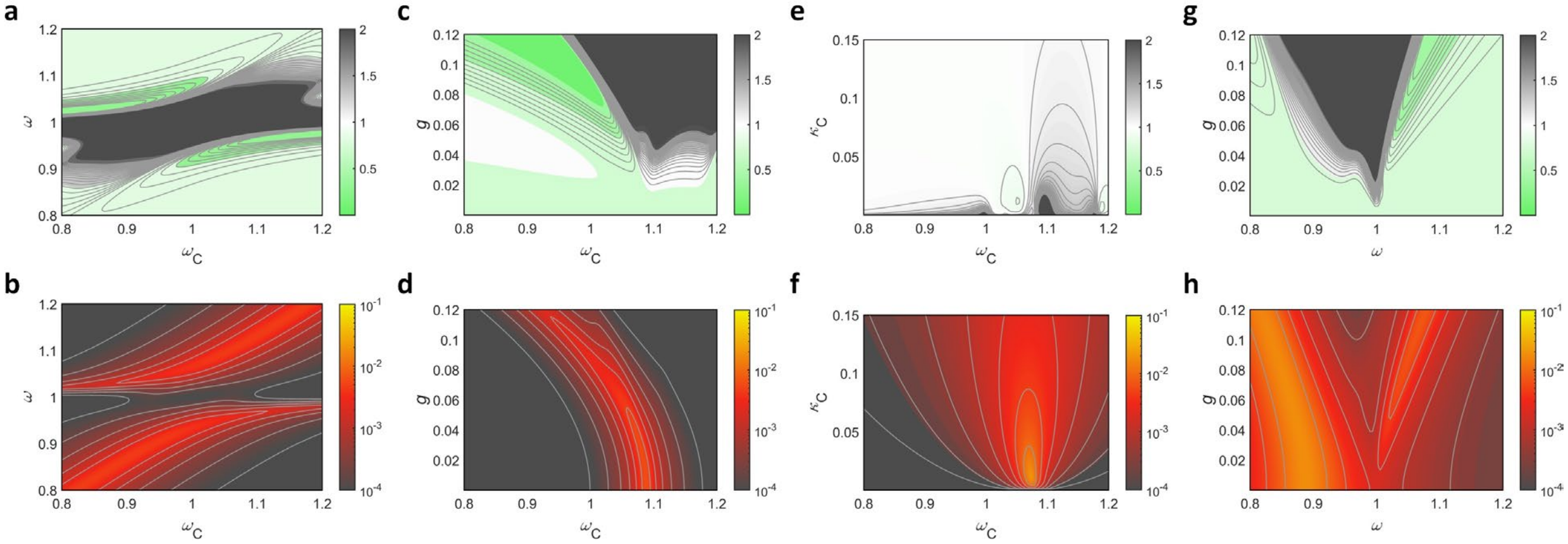

**Fig. S1. Landscapes across design parameters**. **a-h,** Landscapes of $g_2(0)$ (**a,c,e,g**) and $B$ (**b,d,f,h**) over ($\omega_C$, $\omega$) at [$g$, $\xi$, $\kappa_C$] = [0.066, $8.7 \times 10^{-3}$, 0.063] (**a,b**), ($\omega_C$, $g$) at [$\omega$, $\xi$, $\kappa_C$] = [1.088, $6 \times 10^{-3}$, 0.022] (**c,d**), ($\omega_C$, $\kappa_C$) at [$\omega$, $g$, $\xi$] = [1.083, 0.035, 0.010] (**e,f**), and ($\omega$, $g$) at [$\omega_C$, $\xi$, $\kappa_C$] = [0.889, 0.018, 0.073] (**g,h**).

## Supplementary Note S3. Threshold dependence

We examine the dependence of optimization performance on the target threshold $g_{2,\mathrm{th}}$. Figure S2 shows the initial and final optimized sources for various values of $g_{2,\mathrm{th}}$. For stringent thresholds of small $g_{2,\mathrm{th}}$, successful realizations are concentrated in the low $g_2(0)$ region with finite brightness, whereas many failed realizations remain trapped near the coherent plateau similar to Fig. 3a of the main text. As the threshold is relaxed, the success probability increases because the target condition becomes accessible without requiring a large departure from the coherent plateau.

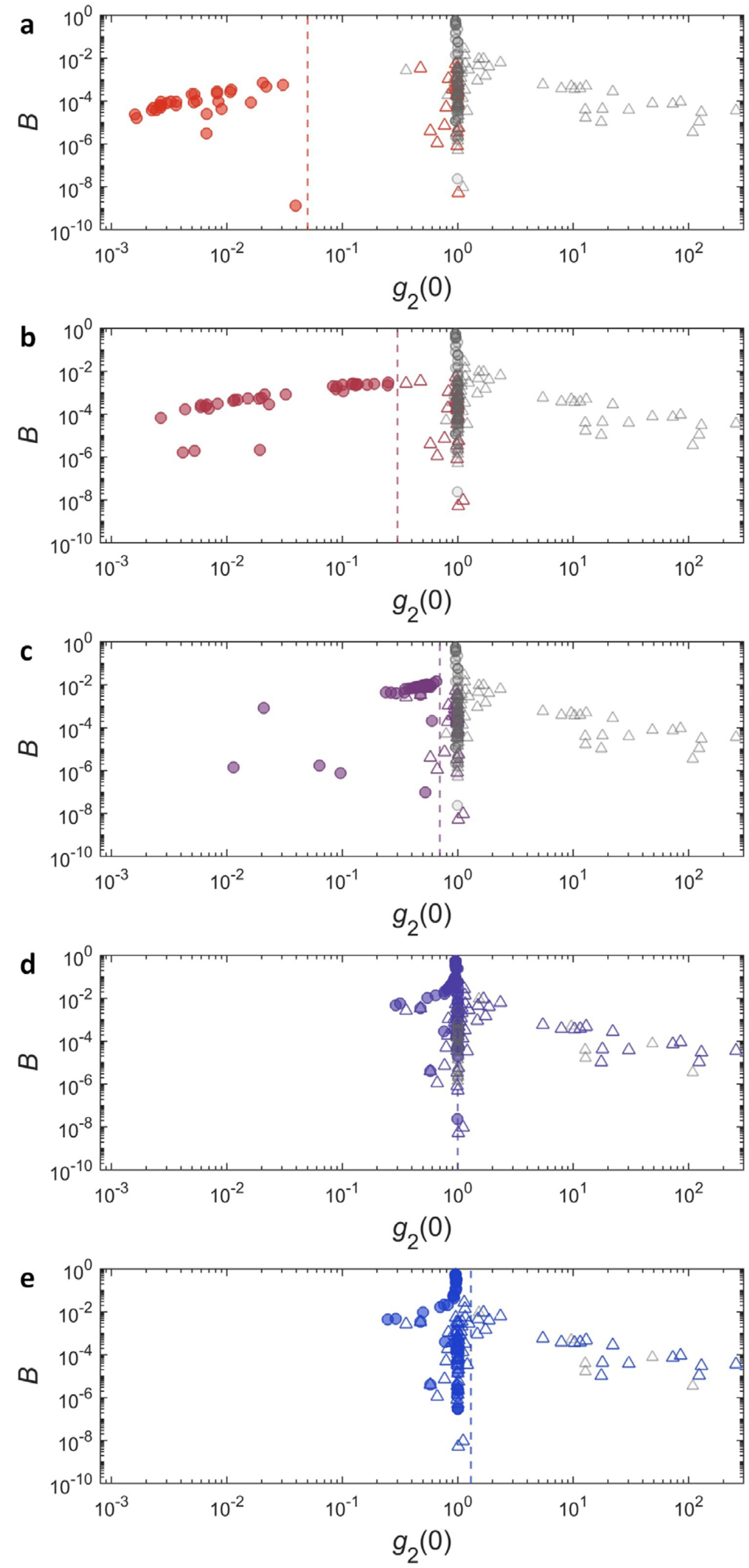

**Fig. S2. Dependence of performance on the threshold**. Initial (triangles) and final (circles) optimized sources for 100 realizations with $g_{2,\mathrm{th}}$ = 0.05 (**a**), 0.3 (**b**), 0.7 (**c**), 1.0 (**d**), and 1.3 (**e**). Coloured and grey markers denote successful and failed realizations, respectively.

Figure S3 further compares the optimization trajectories for successful and failed realizations. Successful cases show a continuous transition toward the target region, followed by the evident brightness recovery (stage II), consistent with the two-stage mechanism described in the main text. In contrast, failed trajectories either stay close to the coherent plateau or permanently lose brightness during the suppression of $g_2(0)$. These results confirm that the coherent plateau is the primary bottleneck limiting optimization.

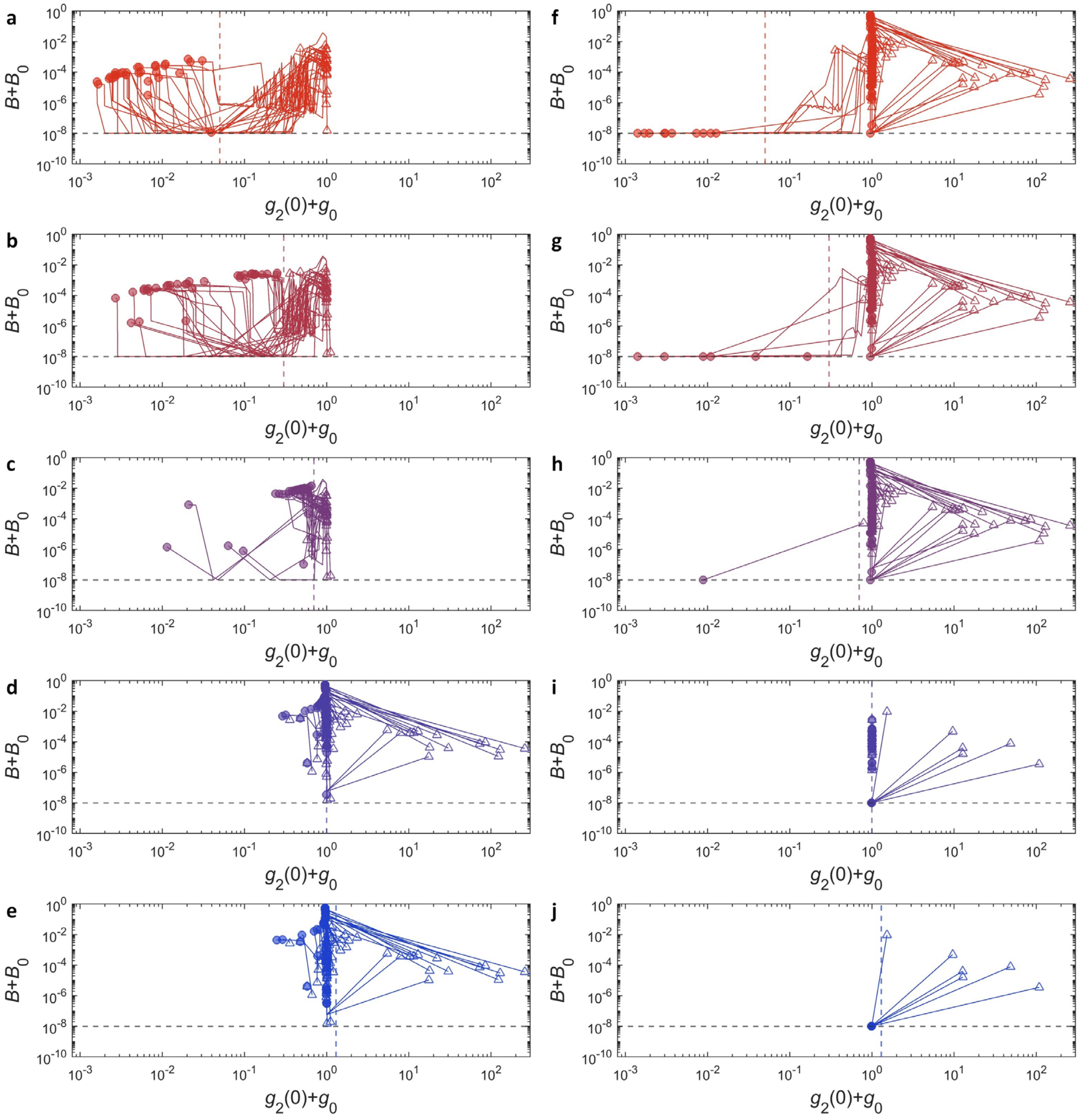

**Fig. S3. Dependence of trajectories on the threshold**. Optimization trajectories for successful (**a-e**) and failed (**f-j**) realizations with $g_{2,\mathrm{th}}$ = 0.05 (**a,f**), 0.3 (**b,g**), 0.7 (**c,h**), 1.0 (**d,i**), and 1.3 (**e,j**).

**Supplementary Note S4. Evolutions of design parameters**

Figure S4 shows the representative evolutions of the design parameters during optimization for $g_{2,\text{th}} = 0.1$ and 0.6. For the stricter threshold ($g_{2,\text{th}} = 0.1$), the first stage is characterized by a rapid suppression of $\kappa_C$ (Fig. S4a) and an increase in $g$ (Fig. 4b), both of which enhance CPB and the consequent decrease of $g_2(0)$. The brightness is recovered in the second stage mainly by increasing $\kappa_C$ (Fig. S4a), while $\xi$ is moderately reduced to avoid degrading the achieved single-photon purity. In contrast, for the relaxed threshold ($g_{2,\text{th}} = 0.6$), the parameter changes are milder, only experiencing the first stage.

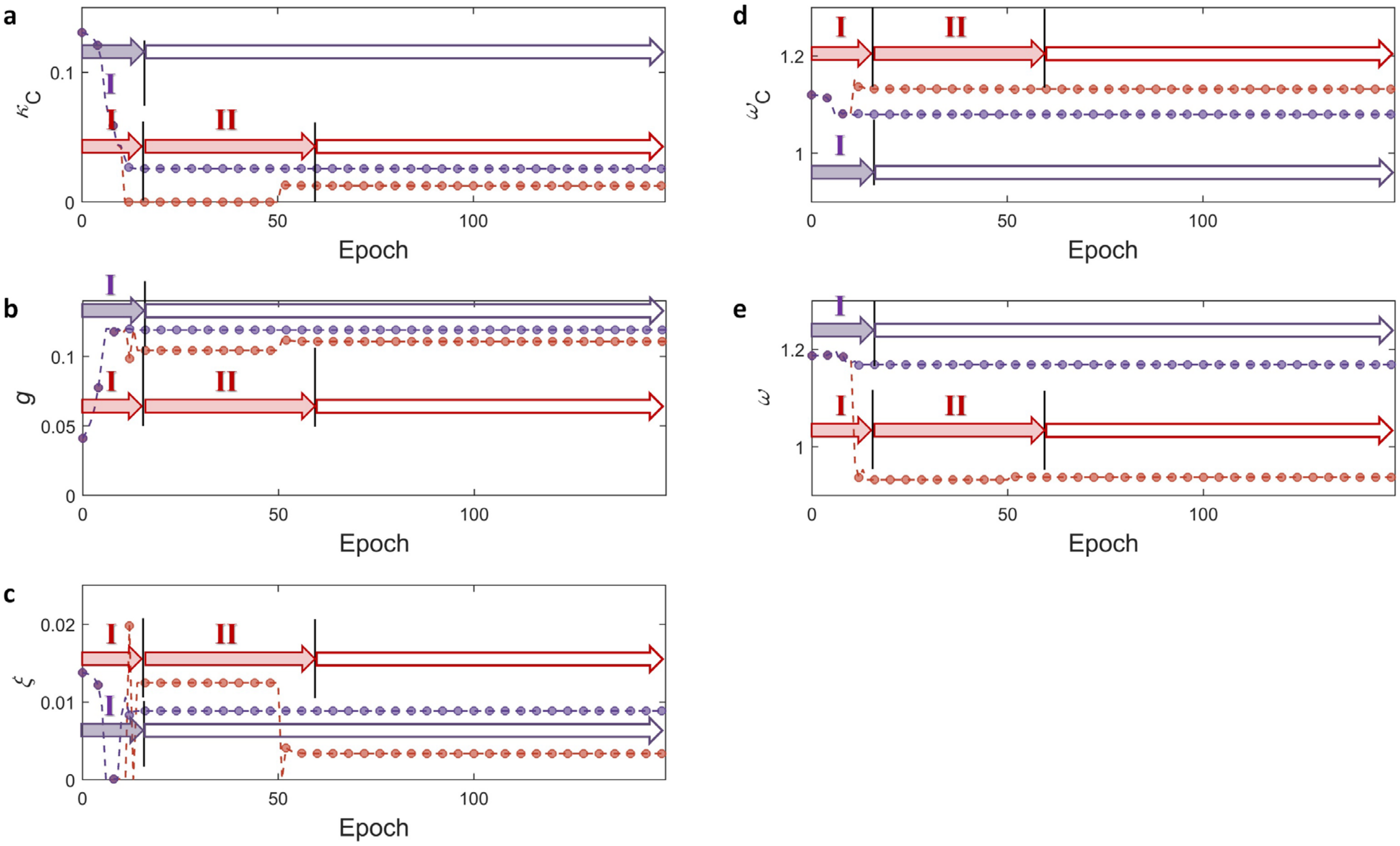

**Fig. S4. Design parameter evolutions**. Evolutions of $\kappa_C$ (**a**), $g$ (**b**), $\xi$ (**c**), $\omega_C$ (**d**), and $\omega$ (**e**), for $g_{2,\text{th}} = 0.1$ (red) and 0.6 (purple).

**Supplementary Note S5. Purity-brightness bound**

To examine the CPB condition in our problem, we start with the traditional undriven and isolated JC model:

$$H_{\mathrm{JC}} = \hbar\omega_{\mathrm{C}} a^\dagger a + \frac{E_{10}}{2}\sigma_z - ig\left(a\sigma_+ - a^\dagger\sigma_-\right), \tag{S7}$$

which leads to the eigenenergies of the coupled states of $\{|g, n\rangle, |e, n-1\rangle\}$, which preserve the total excitation number, for $n \geq 1$:

$$E_n^{\pm} = \left(n - \frac{1}{2}\right)\hbar\omega_{\mathrm{C}} \pm \sqrt{\Delta^2 + ng^2}, \tag{S8}$$

where $\Delta = (E_{10} - \hbar\omega_{\mathrm{C}})/2$, while the ground state $|g, 0\rangle$ supports $E_0 = -E_{10}/2$. We focus on the near-resonance ($\Delta \approx 0$) and strong coupling (large $g$) regime, resulting in the relevant anharmonicity $(E_2^- - E_1^-) - (E_1^- - E_0) = (2 - \sqrt{2})g$, compared with $(E_2^- - E_1^+) - (E_1^+ - E_0) = -(2 + \sqrt{2})g$.

According to Eq. (S8), we apply the driving frequency $\hbar\omega = \hbar\omega_{\mathrm{C}} - g$. From Eqs. (1) and (3) of the main text, the effective Hamiltonian for the driven open JC model near the resonance $\Delta \approx 0$, neglecting quantum jumps and shifting the ground-state energy by $g/2$, becomes

$$H_{\mathrm{eff}} = \left[g - \frac{i\hbar}{2}\left(\kappa_{\mathrm{C}} + \kappa_{\mathrm{L}}\right)\right]a^\dagger a + \left(g - \frac{i\hbar}{2}\gamma\right)\sigma_+\sigma_- - \frac{i\hbar}{4}\gamma_\varphi I - ig\left(a\sigma_+ - a^\dagger\sigma_-\right) + \xi\left(a + a^\dagger\right). \tag{S9}$$

Under the weak-driving limit, we truncate the Hilbert space up to the total excitation number of 2, employing the basis of $\{|g, 0\rangle, |g, 1\rangle, |e, 0\rangle, |g, 2\rangle, |e, 1\rangle\}$. The steady state $H_{\mathrm{eff}}|\psi_{\mathrm{S}}\rangle = \mathbf{0}$ can then be expressed as $|\psi_{\mathrm{S}}\rangle = c_0|g, 0\rangle + c_1^g|g, 1\rangle + c_1^e|e, 0\rangle + c_2^g|g, 2\rangle + c_2^e|e, 1\rangle$. By setting $g_\kappa \equiv g - i\hbar(\kappa_{\mathrm{C}} + \kappa_{\mathrm{L}})/2$ and $g_\gamma \equiv g - i\hbar\gamma/2$, we approximate the no-jump steady-state wavefunction within the truncated basis as

$$
\begin{bmatrix}
-\frac{i\hbar}{4}\gamma_\varphi & \xi & 0 & 0 & 0 \\
\xi & g_\kappa - \frac{i\hbar}{4}\gamma_\varphi & +ig & \sqrt{2}\xi & 0 \\
0 & -ig & g_\gamma - \frac{i\hbar}{4}\gamma_\varphi & 0 & \xi \\
0 & \sqrt{2}\xi & 0 & 2g_\kappa - \frac{i\hbar}{4}\gamma_\varphi & +i\sqrt{2}g \\
0 & 0 & \xi & -i\sqrt{2}g & g_\kappa + g_\gamma - \frac{i\hbar}{4}\gamma_\varphi
\end{bmatrix}
\begin{bmatrix} c_0 \\ c_1^{\ g} \\ c_1^{\ e} \\ c_2^{\ g} \\ c_2^{\ e} \end{bmatrix}
\cong
\begin{bmatrix} 0 \\ 0 \\ 0 \\ 0 \\ 0 \end{bmatrix}. \tag{S10}
$$

Because quantum jumps are neglected, the qubit-dephasing term enters only as a uniform diagonal shift and cancels in the purity and brightness. Setting $c_0 = 1$ and taking $c_1^{g,e} \sim O(\xi)$ and $c_2^{g,e} \sim O(\xi^2)$ under the weak-driving condition $\xi << 1$, we obtain the equations to second order in $\xi$:

$$
\begin{aligned}
& g_\kappa c_1^{\ g} + igc_1^{\ e} = -\xi, \\
& -igc_1^{\ g} + g_\gamma c_1^{\ e} = 0, \\
& 2g_\kappa c_2^{\ g} + i\sqrt{2}gc_2^{\ e} = -\sqrt{2}\xi c_1^{\ g}, \\
& -i\sqrt{2}gc_2^{\ g} + (g_\kappa + g_\gamma)c_2^{\ e} = -\xi c_1^{\ e},
\end{aligned} \tag{S11}
$$

which leads to

$$
\begin{aligned}
c_1^{\ g} &= -\frac{g_\gamma}{g_\kappa g_\gamma - g^2}\xi, \\
c_1^{\ e} &= -\frac{ig}{g_\kappa g_\gamma - g^2}\xi, \\
c_2^{\ g} &= \frac{1}{\sqrt{2}} \frac{g_\gamma (g_\kappa + g_\gamma) + g^2}{g_\kappa (g_\kappa + g_\gamma) - g^2} \frac{1}{g_\kappa g_\gamma - g^2}\xi^2, \\
c_2^{\ e} &= ig\left( \frac{g_\kappa + g_\gamma}{g_\kappa (g_\kappa + g_\gamma) - g^2} \right) \frac{1}{g_\kappa g_\gamma - g^2}\xi^2.
\end{aligned} \tag{S12}
$$

The purity is then estimated as $g_2(0) \approx 2|c_2^g|^2/|c_1^g|^4$, giving

$$g_2(0) = \frac{\left|g_\kappa g_\gamma - g^2\right|^2 \left|g_\gamma (g_\kappa + g_\gamma) + g^2\right|^2}{\left|g_\gamma\right|^4 \left|g_\kappa (g_\kappa + g_\gamma) - g^2\right|^2},. \tag{S13}$$

In the strong coupling regime, the factors reduce to

$$\begin{aligned}
&\left|g_\kappa g_\gamma - g^2\right|^2 \approx \frac{\hbar^2 g^2 \left(\kappa_{\mathrm{C}} + \kappa_{\mathrm{L}} + \gamma\right)^2}{4}, \\
&\left|g_\gamma (g_\kappa + g_\gamma) + g^2\right|^2 \approx 9g^4 \\
&\left|g_\kappa (g_\kappa + g_\gamma) - g^2\right|^2 \approx g^4
\end{aligned} \tag{S14}$$

which lead to

$$g_2(0) \approx \frac{9\hbar^2 \left(\kappa_{\mathrm{C}} + \kappa_{\mathrm{L}} + \gamma\right)^2}{4g^2}. \tag{S15}$$

At the threshold $g_2(0) = g_{2,\mathrm{th}}$, we obtain

$$\kappa_{\mathrm{C}} = \frac{2}{3\hbar} g \sqrt{g_{2,\mathrm{th}}} - \kappa_{\mathrm{L}} - \gamma, \tag{S16}$$

which approximately bounds the outcoupling rate as $\kappa_{\mathrm{C}} \leq (2g/3\hbar) g_{2,\mathrm{th}}^{1/2}$. Because $B = \kappa_{\mathrm{C}} \langle a^\dagger a \rangle$ and the weak-driving condition limits the intracavity photon number to a saturation value $n_{\mathrm{sat}}$, we have $B \leq \kappa_{\mathrm{C}} n_{\mathrm{sat}}$, giving

$$B(g_{2,\mathrm{th}}) \leq \frac{2}{3\hbar} g n_{\mathrm{sat}} \sqrt{g_{2,\mathrm{th}}}. \tag{S17}$$

Equation (S17) determines the brightness bound in the strong-coupling, weak-driving regime. In Figs. 3g and 4b, we set the maximum light–matter coupling energy to $g = 0.12$ and $n_{\mathrm{sat}} = 0.60$.

**Supplementary Note S6. Annealing temperature dependence and evolutions**

Figures S5 and S6 show the dependence of the simulated-annealing-assisted optimization on the initial temperature $T_0$. Across the tested range, successful final designs consistently reach the target region with finite brightness, while the trajectories reveal broader exploration than the deterministic case in Fig. 3b of the main text. In particular, simulated annealing generates additional transition pathways that cross the coherent plateau and access low $g_2(0)$ regions. Figures S7 and S8 further show representative performance and parameter evolutions for $T_0 = 0.04$. These trajectories exhibit a modified second stage, denoted stage II', in which brightness recovery and parameter adjustment proceed differently from the deterministic Jacobian-descent pathway.

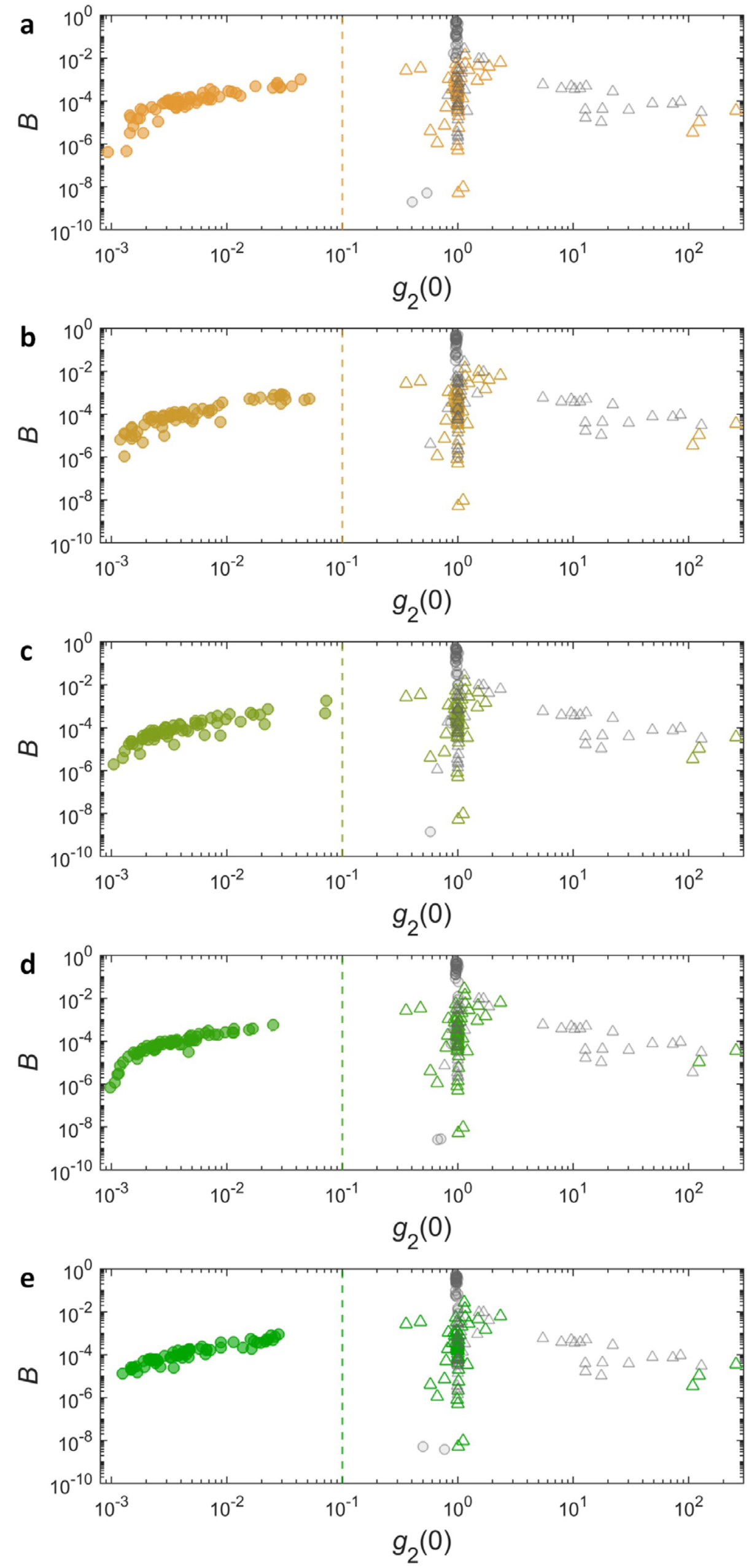

**Fig. S5. Dependence of performance on the temperature**. Initial (triangles) and final (circles) optimized sources for 100 realizations with $T_0 = 0.02$ (**a**), 0.04 (**b**), 0.10 (**c**), 0.16 (**d**), and 0.20 (**e**). Coloured and grey markers denote successful and failed realizations, respectively.

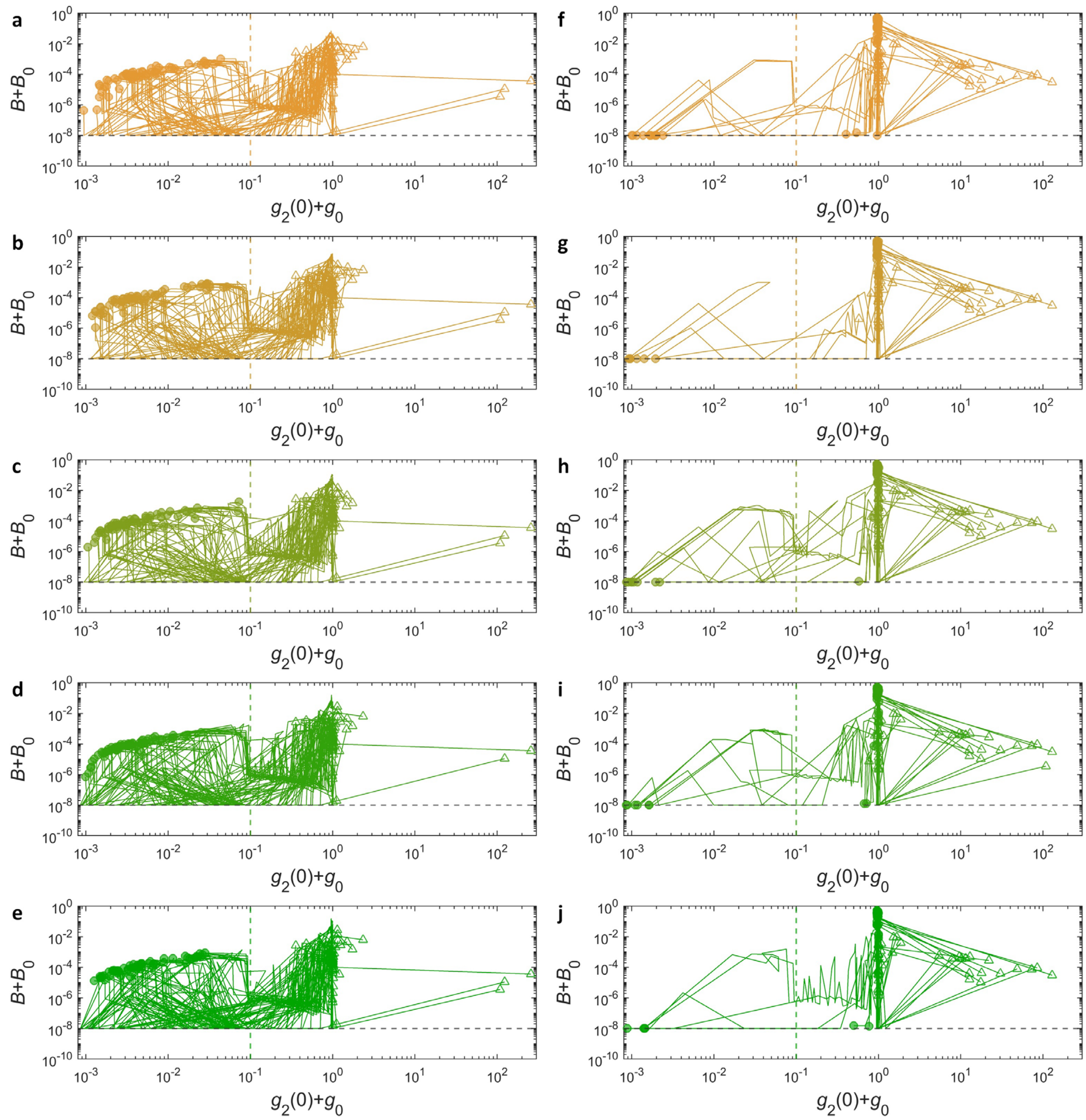

**Fig. S6. Dependence of trajectories on the temperature**. Optimization trajectories for successful (**a-e**) and failed (**f-j**) realizations with $T_0$ = 0.02 (**a,f**), 0.04 (**b,g**), 0.10 (**c,h**), 0.16 (**d,i**), and 0.20 (**e,j**).

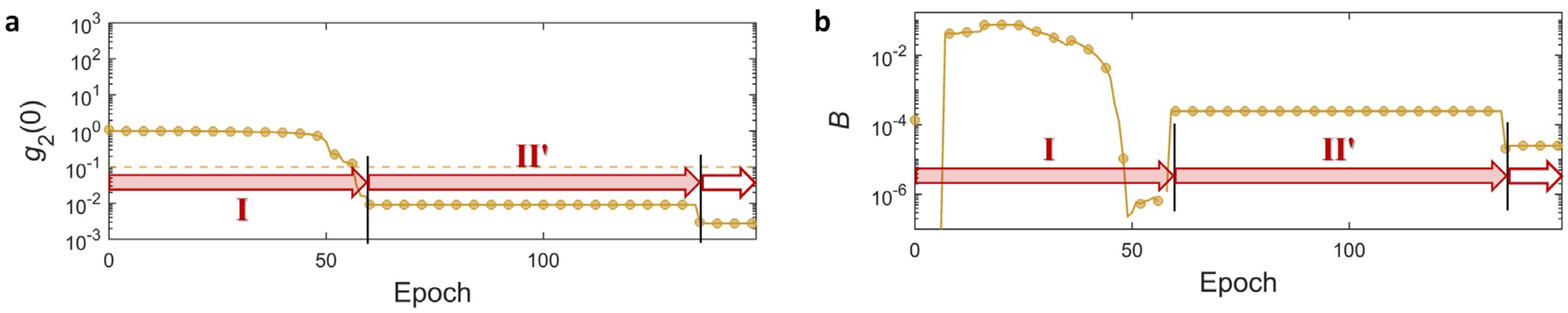

**Fig. S7. Performance evolutions under simulated annealing**. Evolutions of $g_2(0)$ (**a**) and $B$ (**b**) for $T_0 = 0.04$.

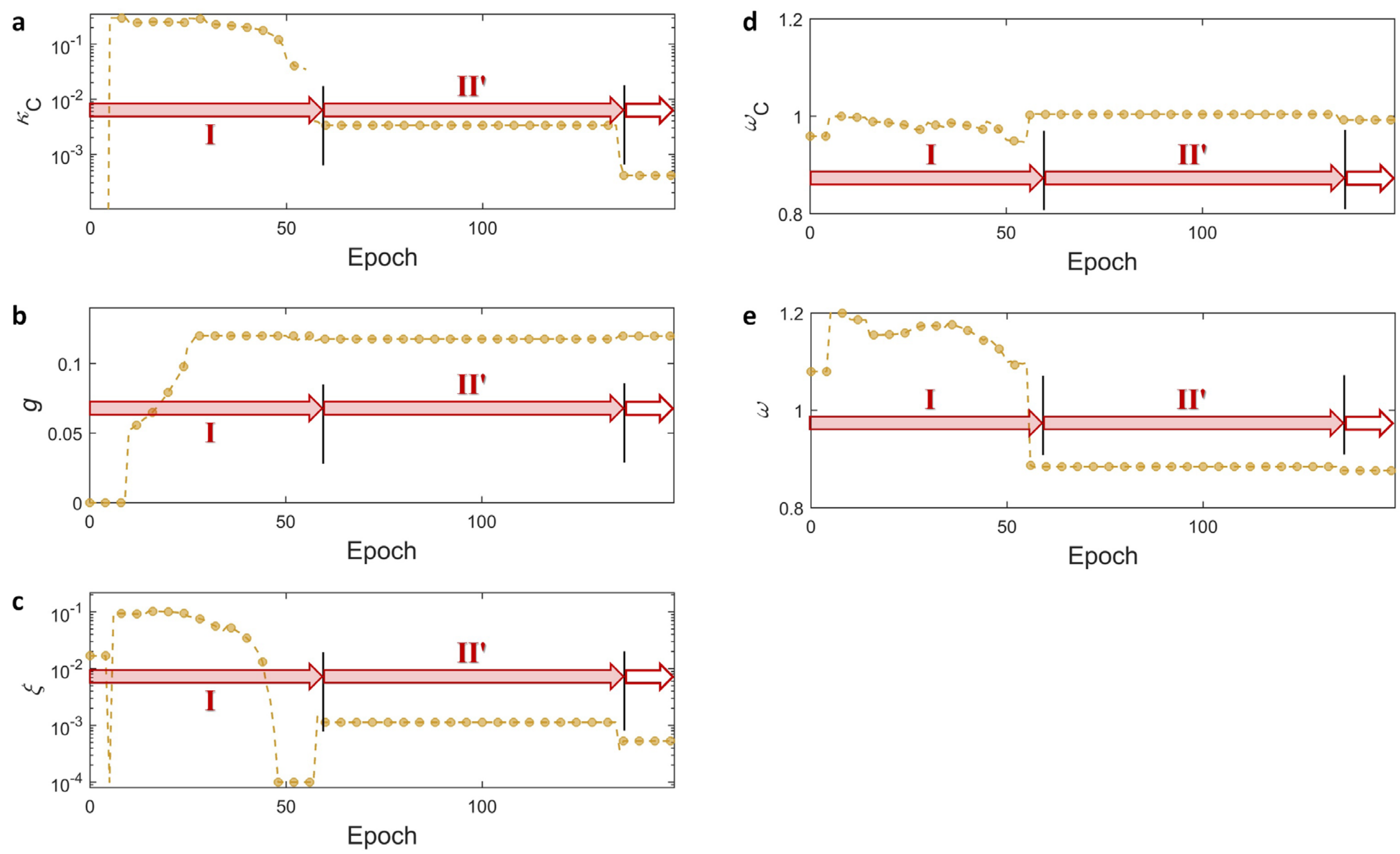

**Fig. S8. Design parameter evolutions under simulated annealing**. Evolutions of $\kappa_C$ (**a**), $g$ (**b**), $\xi$ (**c**), $\omega_C$ (**d**), and $\omega$ (**e**), for $T_0 = 0.04$.

**Supplementary References**

[1] O. Di Stefano, A. Settineri, V. Macrì, L. Garziano, R. Stassi, S. Savasta, and F. Nori, Resolution of gauge ambiguities in ultrastrong-coupling cavity quantum electrodynamics, Nat. Phys. **15**, 803 (2019).